\documentclass[aps,prl,twocolumn,superscriptaddress,nofootinbib,bibnotes,floatfix,longbibliography,preprintnumbers]{revtex4-2}
\usepackage[utf8]{inputenc}
\usepackage[T1]{fontenc}
\usepackage{amsmath,amssymb,amsfonts}
\usepackage{bm}
\usepackage{bbold}
\usepackage{physics}
\usepackage{graphicx}
\usepackage{xcolor}
\usepackage{tikz,tikz-cd}
\usetikzlibrary{arrows.meta,decorations.pathmorphing,calc}
\usepackage{multirow}
\usepackage{booktabs}
\usepackage{comment}
\usepackage{dcolumn}
\usepackage{orcidlink}
\usepackage{hyperref}% add hypertext capabilities
\hypersetup{
	colorlinks=true,
	linkcolor=blue,
	filecolor=magenta,      
	urlcolor=blue,
	citecolor=magenta
}
\newcommand{\mcO}{\mathcal{O}}

\newcommand{\bQ}{\mathbf{Q}}

\newcommand{\ii}{i}

\def\[{\left[}
\def\]{\right]}
\def\({\left(}
\def\){\right)}

\newcommand{\beq}{\begin{eqnarray}}
\newcommand{\eeq}{\end{eqnarray}}
\newcommand{\beqa}{\begin{eqnarray}}
\newcommand{\eeqa}{\end{eqnarray}}

\newcommand{\eq}[1]{(\ref{#1})}

\definecolor{colA}{RGB}{200,30,30}
\definecolor{colB}{RGB}{30,80,200}
\definecolor{colN}{RGB}{110,110,110}
\definecolor{newloops}{rgb}{0.70,0.10,0.10}
\newcommand{\cA}{\mathcal A}
\newcommand{\cB}{\mathcal B}
\newcommand{\cN}{\mathcal N}

\begin{document}

\preprint{QMUL-PH-26-42}

\title{Structure Constants of Planar $\mathcal{N}=4$ SYM at Any Coupling}

\author{Simon Ekhammar\,\orcidlink{0000-0003-1083-6437}}
\email{simon.ekhammar@kcl.ac.uk}
\affiliation{Department of Mathematics, King’s College London, Strand WC2R 2LS, London, United Kingdom}

\author{Alessandro Georgoudis\,\orcidlink{0000-0002-1601-4218}}
\email{Alessandro.georgoudis@city.ac.uk}
\affiliation{Department of Mathematics, City St George’s,
University of London, Northampton Square, EC1V 0HB, London, UK}
\affiliation{School of Physical and Chemical Sciences, Queen Mary University of London, London E1 4NS, United Kingdom}

\author{Nikolay Gromov\,\orcidlink{0000-0002-9124-5526}}
\email{nikolay.gromov@kcl.ac.uk}
\affiliation{Department of Mathematics, King’s College London, Strand WC2R 2LS, London, United Kingdom}

\date{\today}

\begin{abstract}
We recast a recent conjecture for structure constants in planar $\mathcal{N}=4$ supersymmetric Yang--Mills theory,
in a separation-of-variables-inspired form suitable for non-perturbative analysis. All state-dependent information enters through the quantum spectral curve. As a showcase, we compute at finite coupling the three-point functions of two half-BPS operators and the spin-two member of the Konishi multiplet. Our results span weak to strong coupling, reproduce the known perturbative and string-theory limits, and are consistent with conformal-bootstrap bounds. We also obtain new six- and seven-loop perturbative coefficients and extract new strong-coupling data from high-precision numerical fits.
\end{abstract}

\maketitle

%% ---------- Main text -------------------------------------------------
%% REVTeX \section headings; PRL typesets them as unnumbered run-in heads.

\section{Introduction}
Exactly solving an interacting quantum field theory in four dimensions remains a central challenge. For a conformal theory, this means determining the scaling dimensions and structure constants of all local operators, which together fix every correlation function. Planar $\mathcal{N}=4$ supersymmetric Yang--Mills (SYM) theory is the leading candidate for the first such solution. Integrability has already provided non-perturbative access to its spectrum through the Quantum Spectral Curve (QSC)~\cite{Gromov:2013pga,Gromov:2014caa}. Here we demonstrate that integrability also yields full non-perturbative control over a set of non-protected structure constants at finite coupling, smoothly interpolating between known weak- and strong-coupling results.

\begin{figure}[t]
  \centering
  \includegraphics[width=\columnwidth]{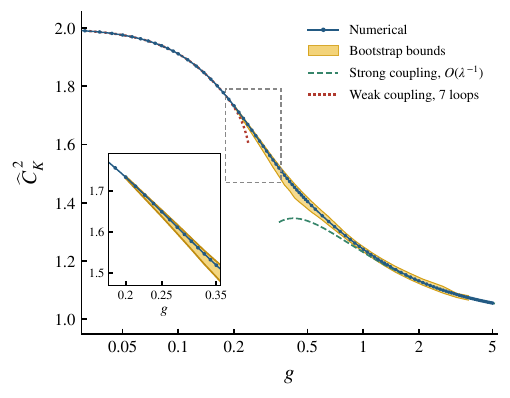}
  \caption{Reduced Konishi structure constant $\widehat C_K^{\,2}=
    \left.\frac{1}{\Delta-1}\left[\frac{\Delta(\Delta^2-4)\binom{\Delta}{\Delta/2}\binom{2p}{p+\Delta/2}}{4p(p-1)\binom{2p}{p}}C_p\right]^2\right|_{p=2}\,,
    $
  \cite{Caron-Huot:2024tzr}; the binomial cancels the double-trace poles. Blue: numerical results; the tree-level value is
  $\widehat C_K^{\,2}=2$. Gold band: bootstrap bounds for $g\geq0.2$ \cite{Caron-Huot:2024tzr}.
  Dashed green: strong-coupling expansion to $O(\lambda^{-1})$
  \cite{Alday:2022uxp,Alday:2022xwz,Alday:2023mvu}. Dotted red: weak-coupling expansion through
  seven loops.}
  \label{fig:konishi}
\end{figure}

We consider the structure constant $C_p$ of two half-BPS operators of protected dimension $p$ with the spin-two member $\mcO_K$ of the Konishi multiplet, whose scaling dimension $\Delta$ depends on the coupling. Schematically,
\begin{equation}
    C_p=\frac{\langle\text{tr}\,Z_{12}D_+^2Z_{12}\big|_{x=0}\,\,\text{tr}\,\bar Z_{12}^p\big|_{x=\infty}\,\,\text{tr}\,Z_{13}^p|_{x=1}\rangle}
    {\langle\text{tr}\,Z_{12}^2|_{x=0}\,\,\text{tr}\,\bar Z_{12}^p|_{x=\infty}\,\,\text{tr}\,Z_{13}^p|_{x=1}\rangle}\,,
    \label{eq:3pt}
\end{equation}
where $\Phi_{i}$ are the real scalars of the theory and $Z_{ij}=\Phi_i+i\Phi_j$. All operators have unit two-point functions, normalised so that $C_2^2=1/6$ at tree level, and $g=\sqrt\lambda/4\pi$ with $\lambda$ the 't Hooft coupling.

Structure constants have been studied intensively over the last fifteen years using integrability. They have been treated at tree level and one loop using tailoring techniques~\cite{Escobedo:2010xs,Gromov:2012uv,Vieira:2013wya} and to the first few orders in perturbation theory using hexagons~\cite{Basso:2015zoa,Basso:2015eqa,Eden:2015ija,Basso:2017muf}. For a particular class of correlators, these developments led to a non-perturbative conjecture for $C_p$ \cite{Basso:2022nny}, which serves as our starting point. It had not, however, been feasible to evaluate it at finite coupling. Three new results make this possible: a determinant representation of one of its factors in terms of Q-functions with a prescription for the analytic continuation in $\Delta$; a closed relation between two of its building blocks that were previously treated as independent; and an algorithm that evaluates its normalisation directly from the QSC solution. They are combined with a high-precision QSC solver for the Q-functions~\cite{Gromov:2015wca,Gromov:2023hzc,Ekhammar:2024rfj} and our optimised C++ implementation of numerical Fredholm-determinant methods~\cite{Bornemann:2010}.

Our result for $C_p$ interpolates from weak to strong coupling, and we find agreement with weak-coupling expansions from four-point functions \cite{Eden:2012rr,Eden:2016aqo,Goncalves:2016vir,Georgoudis:2017meq,Chicherin:2018avq} and strong-coupling predictions from string theory \cite{Bargheer:2013faa,Goncalves:2014ffa,Alday:2022uxp,Alday:2023flc,Alday:2022xwz,Alday:2023mvu}. Furthermore, our results satisfy the bootstrap bounds sharpened by QSC data and supersymmetric localisation \cite{Beem:2013qxa,Beem:2016wfs,Caron-Huot:2022sdy,Caron-Huot:2024tzr}. For example, at $p=2$, Figure~\ref{fig:konishi} shows our results as a smooth curve obtained by removing the simple poles at the double-trace dimensions $\Delta=2p+2n$ with $n\geq1$, corresponding to composites of two half-BPS operators of dimension $p$ (see Supplemental Material for $p=2,3,4$).

\section{Magnons, hexagons and Q-functions}

At weak coupling, integrability has an intuitive quasi-particle picture~\cite{Minahan:2002ve,Beisert:2003yb}: an operator $\text{tr}\,\nabla^{S}Z^{L}$ is a state of $S$ magnons on top of the BPS vacuum $\text{tr}\,Z^L$. The magnon momenta are encoded in rapidities $u_j$, which are quantised by quasi-periodicity on the length-$L$ chain. Scattering is described by an S-matrix fixed by symmetry up to a scalar factor~\cite{Beisert:2005tm}, which is constrained by crossing symmetry~\cite{Janik:2006dc}. In the $\mathfrak{sl}_2$ sector these quantisation conditions take the form of the Beisert--Staudacher equations~\cite{Beisert:2005fw}, $e^{\ii\Phi(u_j)}=1$, with
\begin{equation}
    e^{\ii\Phi(u_j)}\equiv
    \left(\frac{x_j^+}{x_j^-}\right)^{\!L}
    \prod_{k\neq j}^{S}
    \frac{x_j^+-x_k^-}{x_j^--x_k^+}\,
    \frac{1-1/(x_j^-x_k^+)}{1-1/(x_j^+x_k^-)}\,
    \sigma^{-2}(u_j,u_k)\,,
    \label{eq:SpectrumTree}
\end{equation}
where $x_j^\pm=x(u_j\pm\tfrac{\ii}{2})$, $x+1/x=u/g$, and $\sigma$ is the Beisert--Eden--Staudacher (BES) dressing phase~\cite{Beisert:2006ez}. This asymptotic description holds until wrapping corrections set in, which are delayed to order $g^{2L+4}$~\cite{Kotikov:2007cy,Bajnok:2008bm}. 

This picture also guides the structure constants: the chain of the non-trivial operator is split into two subchains as it transitions into the pair of BPS operators, distributing the magnons between them~\cite{Escobedo:2010xs,Basso:2015zoa}. Schematically,
\begin{equation}
    C^{\rm asym}\;\propto\;
    \frac{\sum_{\rm partitions}w_{\alpha}}{\sqrt{\det 
{\cal G}}}\,,
    \qquad {\cal G}_{jk}=\frac{\partial\Phi(u_j)}{\partial u_k}\,,
    \label{eq:StructureTree}
\end{equation}
where $w_\alpha$ is the weight of each partition $\alpha$, and the denominator accounts for the Gaudin norm.

At finite coupling, virtual particles wrap the chain~\cite{Bajnok:2008bm}, obliterating the magnon picture but not integrability. These wrapping effects are resummed by the thermodynamic Bethe ansatz~\cite{Bombardelli:2009ns,Gromov:2009tv,Arutyunov:2009ur}, from which the QSC, the workhorse of our construction, was derived~\cite{Gromov:2013pga,Gromov:2014caa}. The Baxter Q-function provides the bridge to finite coupling: at weak coupling it is simply the polynomial $Q(u)=\prod_{j=1}^S(u-u_j)$, whose zeros are the magnon rapidities. The QSC generalises this polynomial to a finite set of Q-functions, related by the Q-system and characterised by Zhukovsky square-root cuts at $u\in[-2g,2g]+\ii\mathbb Z$. The spectrum is fixed by gluing conditions across the cuts together with prescribed large-$u$ asymptotics, which encode the scaling dimension $\Delta$. The QSC provides high-precision numerical access to the spectrum at finite coupling~\cite{Gromov:2015wca,Gromov:2023hzc}, systematic weak-coupling expansions~\cite{Marboe:2014gma,Marboe:2018ugv}, and strong-coupling data~\cite{Gromov:2014bva}.

Of the $2^8=256$ Q-functions of the QSC, our construction requires only the four fundamental functions $\bQ_i$, $i=1,\dots,4$. They are analytic in the upper half-plane, with large-$u$ asymptotics encoding the dimension $\Delta$ and spin $S$. Their analytic continuations $\widetilde{\bQ}_i$ through the cut $[-2g,2g]$ are related to their complex conjugates by a constant gluing matrix:
\begin{equation}
    \bQ_{i} \simeq B_{i}\,u^{M_{i}}\,,
    \qquad
    \widetilde{\bQ}_{i}(u) = G_{i}{}^{j}\,\overline{\bQ}_{j}(u)\,,
    \label{eq:AsymptoticsQ}
\end{equation}
with $M_{i} = \frac{1}{2}\{\Delta-S,\Delta+S-2,-\Delta-S,-\Delta+S-2\}$ and a constant gluing matrix $G$. For the states considered here its only non-zero entries are $G_1{}^3,G_2{}^4,G_3{}^1,G_4{}^2$, so that $\widetilde\bQ_{1,2,3,4}\sim  u^{M_{3,4,1,2}}$. The gluing conditions are the imprint of unitarity.

\paragraph{Meaning of Q-functions.}
Although the magnon picture is lost, the Q-functions keep a physical meaning: they are the building blocks of the wave function in the basis of separated variables (SoV)~\cite{Sklyanin:1995bm,Gromov:2016itr,Ryan:2018fyo}, in which scalar products become determinants of Q-bilinears~\cite{Cavaglia:2019pow,Gromov:2020fwh,Gromov:2022waj}. In AdS/CFT this picture has produced Wilson-line structure constants as Q-overlaps~\cite{Cavaglia:2018lxi,Giombi:2018qox}, and SoV correlators in fishnet theories~\cite{Cavaglia:2021mft,Upcoming2D} and at weak coupling~\cite{Bercini:2022jxo,Bargheer:2026kon}. These results suggest that structure constants should admit a representation as determinants of integrals of Q-functions.

\paragraph{Structure constants as SoV overlaps.}
Building on the hexagon formalism~\cite{Basso:2015zoa,Basso:2015eqa,Eden:2015ija,Basso:2017muf}, Basso, Georgoudis and Klemenchuk Sueiro conjectured~\cite{Basso:2022nny} that at finite coupling the structure constant is a product of three factors,
\begin{equation}
    C_p=\cN\,\cA\,\cB_p\,.
    \label{eq:NAB}
\end{equation}
The normalisation $\cN$ depends on the Konishi state alone and is the finite-coupling analogue of the Gaudin norm in~\eqref{eq:StructureTree}. In the hexagon picture the three operators are pairwise connected by bridges of propagators; for \eqref{eq:3pt} the two bridges adjacent to the Konishi operator have length one and the bottom bridge, between the BPS operators, has length $p-1$. The adjacent factor $\cA$ and the $p$-dependent bottom factor $\cB_p$ are series over virtual mirror excitations crossing, respectively, the adjacent bridges and the bottom bridge. This conjecture reproduces the known five-loop result at weak coupling~\cite{Georgoudis:2017meq} and the string limit at strong coupling~\cite{Basso:2022nny,Basso:2025mca}.

We managed to recast this conjecture into a form that is genuinely advantageous at finite $g$. Let us sketch the derivation. The {\it bottom} was initially defined as a series of multiple integrals, with one-particle weights given by transfer matrices and pairwise interactions given by the hexagon measure.
As observed in~\cite{Jiang:2016ulr,Basso:2017muf}, the measure factorises into elementary factors $(x-y)/(xy-1)$, one for each pair of the magnons' Zhukovsky variables $x^{[\pm a]}$, $y^{[\pm b]}$. This allows the $N$-magnon integrand to be written as a single $2N\times 2N$ Pfaffian, which turns the series into a Fredholm Pfaffian.
Expanding the elementary factor in inverse powers of $x$ and $y$ reduces the latter to the Pfaffian of a semi-infinite matrix. For parity-even states, such as Konishi,
the matrix becomes block off-diagonal, and the Pfaffian collapses to a determinant.
The conjecture of~\cite{Basso:2022nny} upgrades the asymptotic transfer matrices to their finite-length counterparts, expressed through Q-functions of the QSC. Combining these two observations, we can recast the {\it bottom} as follows:
\begin{equation}
    \cB_p=\det_{m,n\geq1}\bigl[\mathbf 1_{m\leq n}+\mathcal H^{(p)}_{mn}\bigr],
    \label{eq:Bdet}
\end{equation}
where $\mathbf 1_{m\leq n}$ is the upper-triangular matrix of ones and the moments are
\begin{equation}
    \mathcal H^{(p)}_{mn}=\frac{2i}{g}\sum_{a\geq1}\int\!\frac{du}{2\pi}\,\chi^{ij}\,
    \mathbf q_{2m-2,i}^{[a]}\,
    \widetilde{\mathbf q}_{2n-1,j}^{[-a]}\,,
    \label{eq:Bmom}
\end{equation}
with the weighted Q-functions $\{\mathbf q_{n,i},\widetilde{\mathbf q}_{n,i}\}\equiv x^{-p-n}(x-1/x)^{-1}\{\bQ_i,\widetilde\bQ_i\}$, the shift $f^{[\pm a]}(u)=f(u\pm\tfrac{ia}{2})$ with $a$ the mirror bound-state number, the integral running along the real axis, and the Q-system symplectic form, $\bQ^i=\chi^{ij}\bQ_j$, with only nonzero components $\chi^{41}=\chi^{23}=-\chi^{14}=-\chi^{32}=1$. We also verified the new representation \eq{eq:Bdet} independently: at weak coupling to $5$ loops for $p=2$, and at strong coupling we reproduced the first two nontrivial coefficients.

The determinant representation in Eq.~\eqref{eq:Bdet} suggests an SoV interpretation as an overlap between the Konishi state and the vacuum: the Q-functions in the moments carry all the state dependence, while the powers of $1/x$ and the triangular matrix are fixed vacuum data. Similar determinant structures appear in the SoV formalism for 2D fishnets and at large spin~\cite{Upcoming2D,upcomingLS}.

The adjacent factor involves the physical magnons explicitly and appears more complicated. Crossing relations nevertheless suggest a close connection with the bottom contribution, as demonstrated in the strong-coupling BMN regime~\cite{Basso:2019diw} and at leading order in small spin~\cite{Basso:2025mca}. For Konishi, we computed $\cA$ independently from its original definition~\cite{Basso:2022nny}, both numerically at finite coupling and analytically to high order at weak coupling. Comparing it with the bottom contribution, we found the relation
\begin{equation}
    \cA = g^{2}\,|\hat\alpha|\,\frac{\Gamma\left(\frac{\Delta-2}{2}\right)}{\Gamma\left(\frac{6-\Delta}{2}\right)}\,\cB_{2}\,,
    \label{eq:AfromB}
\end{equation}
where $\hat\alpha=B_1B_2\,G_4{}^2$ combines the gluing entry with the leading coefficients of \eqref{eq:AsymptoticsQ}; its modulus is gauge invariant. We checked it analytically through five loops, and numerically at nine couplings between $g=0.1$ and $g=5$, where it holds to $10^{-21}$ at $g=0.1$, better than $10^{-15}$ for $g\leq1.5$, and $10^{-8}$ at $g=5$; in each case the difference is within the numerical uncertainty of our evaluation. We use it to evaluate $\cA$ in what follows.

\paragraph{Normalisation.}
The normalisation is the only factor for which we do not yet have a covariant QSC representation. We instead use the thermodynamic Bethe ansatz, whose coupled integral equations describe wrapping effects through functions $Y_{a,s}(u)$. Their indices form a T-shaped lattice, the T-hook. The functions $Y_{a,0}$ describe the mirror bound states labelled by $a=1,2,\ldots$ in Eq.~\eqref{eq:Bmom}; the auxiliary functions with $s\ne0$ describe their internal degrees of freedom and form the two wings. All these functions are reconstructed from the QSC solution. The proposal of Ref.~\cite{Basso:2022nny} reads
\begin{equation}
    \cN^2=e^{-\mathcal W}\,\frac{\mathcal D_{\rm L}\mathcal D_{\rm R}}{\mathcal D}\,.
    \label{eq:Nnorm}
\end{equation}
Here $\mathcal D$ is the Fredholm determinant obtained by linearising the TBA equations, and $\mathcal D_{\rm L,R}$ are its minors restricted to the two wings. This generalises the Gaudin construction in Eq.~\eqref{eq:StructureTree} from finitely many rapidities to infinitely many functions. The integration contours also include the contributions of the magnon rapidities $u_j$. The Gaussian contribution $\mathcal W$ is defined by the mirror-particle interaction integrals of Ref.~\cite{Basso:2022nny}.

\section{Numerical algorithm}
Starting from QSC solutions accurate to more than 100 digits, we evaluate the determinants in Eqs.~\eqref{eq:Bdet} and \eqref{eq:Nnorm} to obtain the structure constant $C_p$.

\paragraph{Bottom.} We evaluate the infinite determinant by truncating it to a finite matrix. This is justified because increasing either index of $\mathcal H^{(p)}_{mn}$ in Eq.~\eqref{eq:Bmom} introduces additional inverse powers of the shifted Zhukovsky variables, whose moduli exceed one on the integration contours, suppressing higher moments.

Some low-index moments, however, become divergent as $g$ increases because of the large-rapidity growth of the Q-functions in \eqref{eq:AsymptoticsQ}.
We define these integrals by analytic continuation in $\Delta$ from their domain of convergence; related continuations appear in \cite{Cavaglia:2021mft,Upcoming2D}. Genuine poles appear at resonant dimensions. Away from these poles, we subtract the large-$u$ asymptotics and add back their analytically continued integrals.

Subtracting sufficiently many asymptotic terms leaves rapidly decaying remainders. The bound-state sum then gives a double integral with a digamma kernel for the remainder--remainder term, single integrals with Hurwitz zeta kernels for the mixed terms, and explicit zeta functions for the asymptotic--asymptotic term. We evaluate the remaining integrals by quadrature. The zeta functions expose possible poles at the double-trace dimensions $\Delta=2p+2,2p+4,\ldots$; numerically, we find only simple poles at the dimensions probed.

\paragraph{Normalisation.}
We simplify the determinant ratio in Eq.~\eqref{eq:Nnorm} using the nearest-neighbour couplings of the simplified TBA to eliminate the auxiliary variations, leaving variations of $Y_{a,0}$ and of the magnon rapidities $u_j$. Variations of $Y_{a,0}$ give a Fredholm determinant, while variations of $u_j$ give a Gaudin determinant generalising Eq.~\eqref{eq:StructureTree} to include wrapping effects. The Gaussian contribution $\mathcal W$ depends on $Y_{a,0}$, rather than on their variations. The QSC supplies all state-dependent inputs needed to evaluate $\mathcal W$ and construct both determinants, avoiding a separate solution of the nonlinear TBA equations.

For numerical evaluation, we retain a finite number of mirror bound states and estimate the remaining contribution by extrapolation. We then use the Nystr\"om method~\cite{Bornemann:2010}: replacing integrals by weighted quadrature sums gives a finite matrix whose determinant we compute. Numerical uncertainties are estimated by varying the number of bound states retained, the extrapolation fit range and order, the cutoffs in auxiliary sums and expansions, the quadrature grids, and the working precision.

\section{Results}

\paragraph{Finite coupling.}
We display our numerical results for the Konishi structure constant in Fig.~\ref{fig:konishi}, from weak to strong coupling. The numerical curve reproduces the Feynman-diagram results at weak coupling~\cite{Georgoudis:2017meq,Chicherin:2018avq}, approaches the string-theory prediction at large $g$~\cite{Alday:2022uxp,Alday:2022xwz,Alday:2023mvu}, and lies within the bootstrap bounds at every shared coupling~\cite{Caron-Huot:2024tzr}. We have also computed the bottom contribution for the longer bridges, $p=3,4$; see the Supplemental Material.

\paragraph{Strong coupling.}
Following Refs.~\cite{Caron-Huot:2024tzr,Basso:2025mca}, we consider the reduced structure constant $\widehat C_p$ (as defined in figure~\ref{fig:konishi} for general $p$). The binomial coefficient $\binom{2p}{p+\Delta/2}$ vanishes at the double-trace dimensions $\Delta=2p+2,2p+4,\ldots$ and cancels the simple poles of $\cB_p$. For $p=2$ the first two coefficients follow from the AdS Virasoro--Shapiro amplitude \cite{Alday:2022uxp,Alday:2022xwz,Alday:2023mvu,Caron-Huot:2024tzr} and the hexagon analysis of Ref.~\cite{Basso:2025mca} fixes their $p$ dependence; with the strong-coupling Konishi dimension \cite{Gromov:2011bz,Gromov:2014bva}, $\lambda=(4\pi g)^2$, they give
\begin{equation}
    \log\widehat C_p^{\,2}
    \simeq\frac{\tfrac54+2\zeta_3}{\sqrt\lambda}
    +\frac{\tfrac1{32}-(3+2p^2)\zeta_3-3\zeta_5}{\lambda}
    \,.
    \label{eq:Chatstrong}
\end{equation}
Our data for $\widehat C_2$ are in agreement with \eqref{eq:Chatstrong}.

The bottom factor can be studied much more precisely: we computed it to about twenty digits up to $g=36$ for $p=3,4$ and $g=54$ for $p=2$. Normalising it as $\widehat C_{\cB}^{(p)}=\frac{\sqrt\lambda\,e^{-2\gamma_{\rm E}}\,\Gamma(p-1)\Gamma(p+1)\,\Delta(\Delta^2-4)}{\sqrt{\Delta^2-1}\,\Gamma(p+1-\Delta/2)\Gamma(p+1+\Delta/2)}\,\cB_p$, with $\gamma_{\rm E}$ Euler's constant, so that it tends to one at strong coupling and $\widehat C_p/\widehat C_{\cB}^{(p)}$ is independent of $p$, fits of these data give
\begin{eqnarray}
    &&\log\widehat C_{\cB}^{(p)\,2} = \frac{\tfrac14-\tfrac13\zeta_3}{\sqrt\lambda}
    +\frac{-\tfrac{31}{32}+(\tfrac{127}{12}-2p^2)\zeta_3+\tfrac1{10}\zeta_5}{\lambda}\nonumber\\
    &&\quad\quad+\frac{1}{\lambda^{3/2}}\Bigl[-\tfrac{95}{192}+(p^2-\tfrac{149}{12})\zeta_3\nonumber\\
    &&\quad\quad\hspace{0.8em}+(\tfrac{3p^2}{2}-\tfrac{433}{20})\zeta_5-\tfrac{5}{112}\zeta_7-2\zeta_3^2\Bigr]+O(\lambda^{-2})\,.
    \label{eq:CBstrong}
\end{eqnarray}
The first two orders are known analytically~\cite{Basso:2025mca,Gromov:2011bz,Gromov:2014bva}. The $\lambda^{-3/2}$ term is new: its closed form was identified by an integer-relation search.

\paragraph{Weak coupling.}
At weak coupling, the QSC can be solved analytically order by order \cite{Marboe:2018ugv}, and the same formulas then yield the perturbative expansion of $C_p$ exactly. The moments in \eqref{eq:Bmom} scale as $\mathcal H^{(p)}_{jk}=O(g^{2j+2k+2p-4})$, so only finitely many enter at each order, and the remaining sums and integrals can be performed analytically in the function space of the perturbative QSC, yielding rational numbers and multiple zeta values.

At weak coupling the determinants in \eqref{eq:Nnorm} reduce to the Gaudin determinant of the Bethe--Yang equations, corrected from four loops on by the mirror particles through a trace of the linearised kernel over their fillings and their back-reaction on the Bethe roots.

Our evaluation of the three factors reproduces the five-loop result for $C_2$ \cite{Georgoudis:2017meq,Chicherin:2018avq}, including its wrapping corrections, and extends it by two orders.
At six loops we find
\begin{multline}
    [g^{12}](6C_2^2)=3571200+1101504\zeta^{\rm sv}_3-113616(\zeta^{\rm sv}_3)^2\\
    -4320(\zeta^{\rm sv}_3)^3+\zeta^{\rm sv}_5\bigl(771792-67032\zeta^{\rm sv}_3+101520\zeta^{\rm sv}_5\bigr)\\
    +\zeta^{\rm sv}_7(1175958+189504\zeta^{\rm sv}_3)+1267500\zeta^{\rm sv}_9+1829520\zeta^{\rm sv}_{11}\,.
\nonumber    %\label{eq:C22six}
\end{multline}
At seven loops we find
\begin{multline}
    [g^{14}](6C_2^2)=3\bigl[-8058880+8640\zeta^{\rm sv}_{3,5,3}\\
    +\zeta^{\rm sv}_3\bigl(-8266880+572832\zeta^{\rm sv}_3+3456(\zeta^{\rm sv}_3)^2\bigr)\\
    +\zeta^{\rm sv}_5\bigl(-3867520+792864\zeta^{\rm sv}_3+69984(\zeta^{\rm sv}_3)^2-217200\zeta^{\rm sv}_5\bigr)\\
    +\zeta^{\rm sv}_7\bigl(-4675560+260112\zeta^{\rm sv}_3-1126080\zeta^{\rm sv}_5\bigr)\\
    -\zeta^{\rm sv}_9(6090000+1049760\zeta^{\rm sv}_3)-5768904\zeta^{\rm sv}_{11}\\
    -8833968\zeta^{\rm sv}_{13}\bigr]\,.
    \label{eq:C22seven}
\end{multline}
Here $\zeta^{\rm sv}_n=2\zeta_n$ for odd $n$ and $\zeta^{\rm sv}_{3,5,3}=2[\zeta(3,5,3)-\zeta_3\zeta(3,5)]-10\zeta_3^2\zeta_5$, with $\zeta(s_1,\ldots,s_k)=\sum_{0<n_1<\cdots<n_k}\prod_j n_j^{-s_j}$. That the seven-loop coefficient, the first to involve a multiple zeta value of depth three, is expressible entirely in single-valued multiple zeta values is a non-trivial check of the construction. Fits to our numerical data at small $g$ reproduce the two coefficients to relative precision of about $10^{-8}$ and a few times $10^{-7}$, respectively.

\section{Conclusions and outlook}
Our results open a new exciting chapter in the exploration of $\mathcal N=4$ SYM, making structure constants accessible at finite coupling.
Consistency with all available data and bootstrap bounds, together with the single-valued zeta structure of the perturbative expansion, strengthens confidence in the conjectures underlying our construction. The higher-spin generalisation of \eqref{eq:AfromB} should allow the extension along the leading Regge trajectory. Analytic continuation of the QSC in spin~\cite{Gromov:2015wca} then gives access to the light-ray operators governing the Regge limit. More broadly, we expect the method to apply to correlators of a left-right symmetric state and two half-BPS operators; general interacting states require further work.

Our results also provide further clues towards an SoV formulation of $\mathcal N=4$ SYM at finite coupling: the bottom contribution in Eq.~\eqref{eq:Bdet} resembles an SoV overlap, expressed as a determinant of moments of Q-functions, and the full structure constant is computed from the QSC solution alone. We hope that the method will also carry over to other models with a Quantum Spectral Curve, such as ABJM theory~\cite{Cavaglia:2014exa,Bombardelli:2017vhk} and AdS$_3$/CFT$_2$~\cite{Cavaglia:2021eqr,Ekhammar:2021pys}.

%% ---------- Acknowledgments -----------------------------------------
\begin{acknowledgments}
We thank Benjamin Basso, Carlos Bercini, Andrea Cavagli{\`a}, Gabriel Lefundes, Fedor Levkovich-Maslyuk, Paul Ryan, Didina Serban and Pedro Vieira for discussions on related topics. A.G. is supported by a Royal Society grant URF$\backslash$R1$\backslash$241371. The work of N.G. was supported by the Science and Technology Facilities Council (STFC) under the grant ST/X000753/1. The work of N.G.\ was partially supported by the European Research Council (ERC) under the European Union's Horizon 2020 research and innovation programme (grant agreement No.~865075) EXACTC. This research utilized Queen Mary's Apocrita HPC facility,
supported by QMUL Research IT~\url{http://doi.org/10.5281/zenodo.438045}. 

\paragraph{Use of AI tools.}
AI tools were used in preparing this work. Symbolic and numerical verifications were performed in Mathematica through the Wolfbook interface~\cite{Wolfbook}, which records the computations of both the authors and the AI in shared notebooks, so that each step can be inspected. The authors take full responsibility for the content of the paper.
\end{acknowledgments}

%% ---------- Bibliography ----------------------------------------------
\bibliographystyle{apsrev4-2}
\bibliography{ref}

%apsrev4-2.bst 2019-01-14 (MD) hand-edited version of apsrev4-1.bst
%Control: key (0)
%Control: author (72) initials jnrlst
%Control: editor formatted (1) identically to author
%Control: production of article title (-1) disabled
%Control: page (0) single
%Control: year (1) truncated
%Control: production of eprint (0) enabled
\begin{thebibliography}{65}%
\makeatletter
\providecommand \@ifxundefined [1]{%
 \@ifx{#1\undefined}
}%
\providecommand \@ifnum [1]{%
 \ifnum #1\expandafter \@firstoftwo
 \else \expandafter \@secondoftwo
 \fi
}%
\providecommand \@ifx [1]{%
 \ifx #1\expandafter \@firstoftwo
 \else \expandafter \@secondoftwo
 \fi
}%
\providecommand \natexlab [1]{#1}%
\providecommand \enquote  [1]{``#1''}%
\providecommand \bibnamefont  [1]{#1}%
\providecommand \bibfnamefont [1]{#1}%
\providecommand \citenamefont [1]{#1}%
\providecommand \href@noop [0]{\@secondoftwo}%
\providecommand \href [0]{\begingroup \@sanitize@url \@href}%
\providecommand \@href[1]{\@@startlink{#1}\@@href}%
\providecommand \@@href[1]{\endgroup#1\@@endlink}%
\providecommand \@sanitize@url [0]{\catcode `\\12\catcode `\$12\catcode
  `\&12\catcode `\#12\catcode `\^12\catcode `\_12\catcode `\%12\relax}%
\providecommand \@@startlink[1]{}%
\providecommand \@@endlink[0]{}%
\providecommand \url  [0]{\begingroup\@sanitize@url \@url }%
\providecommand \@url [1]{\endgroup\@href {#1}{\urlprefix }}%
\providecommand \urlprefix  [0]{URL }%
\providecommand \Eprint [0]{\href }%
\providecommand \doibase [0]{https://doi.org/}%
\providecommand \selectlanguage [0]{\@gobble}%
\providecommand \bibinfo  [0]{\@secondoftwo}%
\providecommand \bibfield  [0]{\@secondoftwo}%
\providecommand \translation [1]{[#1]}%
\providecommand \BibitemOpen [0]{}%
\providecommand \bibitemStop [0]{}%
\providecommand \bibitemNoStop [0]{.\EOS\space}%
\providecommand \EOS [0]{\spacefactor3000\relax}%
\providecommand \BibitemShut  [1]{\csname bibitem#1\endcsname}%
\let\auto@bib@innerbib\@empty
%</preamble>
\bibitem [{\citenamefont {Gromov}\ \emph
  {et~al.}(2014{\natexlab{a}})\citenamefont {Gromov}, \citenamefont {Kazakov},
  \citenamefont {Leurent},\ and\ \citenamefont {Volin}}]{Gromov:2013pga}%
  \BibitemOpen
  \bibfield  {author} {\bibinfo {author} {\bibfnamefont {N.}~\bibnamefont
  {Gromov}}, \bibinfo {author} {\bibfnamefont {V.}~\bibnamefont {Kazakov}},
  \bibinfo {author} {\bibfnamefont {S.}~\bibnamefont {Leurent}},\ and\ \bibinfo
  {author} {\bibfnamefont {D.}~\bibnamefont {Volin}},\ }\href
  {https://doi.org/10.1103/PhysRevLett.112.011602} {\bibfield  {journal}
  {\bibinfo  {journal} {Phys. Rev. Lett.}\ }\textbf {\bibinfo {volume} {112}},\
  \bibinfo {pages} {011602} (\bibinfo {year} {2014}{\natexlab{a}})},\ \Eprint
  {https://arxiv.org/abs/1305.1939} {arXiv:1305.1939 [hep-th]} \BibitemShut
  {NoStop}%
\bibitem [{\citenamefont {Gromov}\ \emph {et~al.}(2015)\citenamefont {Gromov},
  \citenamefont {Kazakov}, \citenamefont {Leurent},\ and\ \citenamefont
  {Volin}}]{Gromov:2014caa}%
  \BibitemOpen
  \bibfield  {author} {\bibinfo {author} {\bibfnamefont {N.}~\bibnamefont
  {Gromov}}, \bibinfo {author} {\bibfnamefont {V.}~\bibnamefont {Kazakov}},
  \bibinfo {author} {\bibfnamefont {S.}~\bibnamefont {Leurent}},\ and\ \bibinfo
  {author} {\bibfnamefont {D.}~\bibnamefont {Volin}},\ }\href
  {https://doi.org/10.1007/JHEP09(2015)187} {\bibfield  {journal} {\bibinfo
  {journal} {JHEP}\ }\textbf {\bibinfo {volume} {09}},\ \bibinfo {pages}
  {187}},\ \Eprint {https://arxiv.org/abs/1405.4857} {arXiv:1405.4857 [hep-th]}
  \BibitemShut {NoStop}%
\bibitem [{\citenamefont {Caron-Huot}\ \emph {et~al.}(2025)\citenamefont
  {Caron-Huot}, \citenamefont {Coronado},\ and\ \citenamefont
  {Zahraee}}]{Caron-Huot:2024tzr}%
  \BibitemOpen
  \bibfield  {author} {\bibinfo {author} {\bibfnamefont {S.}~\bibnamefont
  {Caron-Huot}}, \bibinfo {author} {\bibfnamefont {F.}~\bibnamefont
  {Coronado}},\ and\ \bibinfo {author} {\bibfnamefont {Z.}~\bibnamefont
  {Zahraee}},\ }\href {https://doi.org/10.1007/JHEP05(2025)220} {\bibfield
  {journal} {\bibinfo  {journal} {JHEP}\ }\textbf {\bibinfo {volume} {05}},\
  \bibinfo {pages} {220}},\ \Eprint {https://arxiv.org/abs/2412.00249}
  {arXiv:2412.00249 [hep-th]} \BibitemShut {NoStop}%
\bibitem [{\citenamefont {Alday}\ \emph
  {et~al.}(2022{\natexlab{a}})\citenamefont {Alday}, \citenamefont {Hansen},\
  and\ \citenamefont {Silva}}]{Alday:2022uxp}%
  \BibitemOpen
  \bibfield  {author} {\bibinfo {author} {\bibfnamefont {L.~F.}\ \bibnamefont
  {Alday}}, \bibinfo {author} {\bibfnamefont {T.}~\bibnamefont {Hansen}},\ and\
  \bibinfo {author} {\bibfnamefont {J.~A.}\ \bibnamefont {Silva}},\ }\href
  {https://doi.org/10.1007/JHEP10(2022)036} {\bibfield  {journal} {\bibinfo
  {journal} {JHEP}\ }\textbf {\bibinfo {volume} {10}},\ \bibinfo {pages}
  {036}},\ \Eprint {https://arxiv.org/abs/2204.07542} {arXiv:2204.07542
  [hep-th]} \BibitemShut {NoStop}%
\bibitem [{\citenamefont {Alday}\ \emph
  {et~al.}(2022{\natexlab{b}})\citenamefont {Alday}, \citenamefont {Hansen},\
  and\ \citenamefont {Silva}}]{Alday:2022xwz}%
  \BibitemOpen
  \bibfield  {author} {\bibinfo {author} {\bibfnamefont {L.~F.}\ \bibnamefont
  {Alday}}, \bibinfo {author} {\bibfnamefont {T.}~\bibnamefont {Hansen}},\ and\
  \bibinfo {author} {\bibfnamefont {J.~A.}\ \bibnamefont {Silva}},\ }\href
  {https://doi.org/10.1007/JHEP12(2022)010} {\bibfield  {journal} {\bibinfo
  {journal} {JHEP}\ }\textbf {\bibinfo {volume} {12}},\ \bibinfo {pages}
  {010}},\ \Eprint {https://arxiv.org/abs/2209.06223} {arXiv:2209.06223
  [hep-th]} \BibitemShut {NoStop}%
\bibitem [{\citenamefont {Alday}\ and\ \citenamefont
  {Hansen}(2023)}]{Alday:2023mvu}%
  \BibitemOpen
  \bibfield  {author} {\bibinfo {author} {\bibfnamefont {L.~F.}\ \bibnamefont
  {Alday}}\ and\ \bibinfo {author} {\bibfnamefont {T.}~\bibnamefont {Hansen}},\
  }\href {https://doi.org/10.1007/JHEP10(2023)023} {\bibfield  {journal}
  {\bibinfo  {journal} {JHEP}\ }\textbf {\bibinfo {volume} {10}},\ \bibinfo
  {pages} {023}},\ \Eprint {https://arxiv.org/abs/2306.12786} {arXiv:2306.12786
  [hep-th]} \BibitemShut {NoStop}%
\bibitem [{\citenamefont {Escobedo}\ \emph {et~al.}(2011)\citenamefont
  {Escobedo}, \citenamefont {Gromov}, \citenamefont {Sever},\ and\
  \citenamefont {Vieira}}]{Escobedo:2010xs}%
  \BibitemOpen
  \bibfield  {author} {\bibinfo {author} {\bibfnamefont {J.}~\bibnamefont
  {Escobedo}}, \bibinfo {author} {\bibfnamefont {N.}~\bibnamefont {Gromov}},
  \bibinfo {author} {\bibfnamefont {A.}~\bibnamefont {Sever}},\ and\ \bibinfo
  {author} {\bibfnamefont {P.}~\bibnamefont {Vieira}},\ }\href
  {https://doi.org/10.1007/JHEP09(2011)028} {\bibfield  {journal} {\bibinfo
  {journal} {JHEP}\ }\textbf {\bibinfo {volume} {09}},\ \bibinfo {pages}
  {028}},\ \Eprint {https://arxiv.org/abs/1012.2475} {arXiv:1012.2475 [hep-th]}
  \BibitemShut {NoStop}%
\bibitem [{\citenamefont {Gromov}\ and\ \citenamefont
  {Vieira}(2014)}]{Gromov:2012uv}%
  \BibitemOpen
  \bibfield  {author} {\bibinfo {author} {\bibfnamefont {N.}~\bibnamefont
  {Gromov}}\ and\ \bibinfo {author} {\bibfnamefont {P.}~\bibnamefont
  {Vieira}},\ }\href {https://doi.org/10.1007/JHEP04(2014)068} {\bibfield
  {journal} {\bibinfo  {journal} {JHEP}\ }\textbf {\bibinfo {volume} {04}},\
  \bibinfo {pages} {068}},\ \Eprint {https://arxiv.org/abs/1205.5288}
  {arXiv:1205.5288 [hep-th]} \BibitemShut {NoStop}%
\bibitem [{\citenamefont {Vieira}\ and\ \citenamefont
  {Wang}(2014)}]{Vieira:2013wya}%
  \BibitemOpen
  \bibfield  {author} {\bibinfo {author} {\bibfnamefont {P.}~\bibnamefont
  {Vieira}}\ and\ \bibinfo {author} {\bibfnamefont {T.}~\bibnamefont {Wang}},\
  }\href {https://doi.org/10.1007/JHEP10(2014)035} {\bibfield  {journal}
  {\bibinfo  {journal} {JHEP}\ }\textbf {\bibinfo {volume} {10}},\ \bibinfo
  {pages} {035}},\ \Eprint {https://arxiv.org/abs/1311.6404} {arXiv:1311.6404
  [hep-th]} \BibitemShut {NoStop}%
\bibitem [{\citenamefont {Basso}\ \emph {et~al.}(2015)\citenamefont {Basso},
  \citenamefont {Komatsu},\ and\ \citenamefont {Vieira}}]{Basso:2015zoa}%
  \BibitemOpen
  \bibfield  {author} {\bibinfo {author} {\bibfnamefont {B.}~\bibnamefont
  {Basso}}, \bibinfo {author} {\bibfnamefont {S.}~\bibnamefont {Komatsu}},\
  and\ \bibinfo {author} {\bibfnamefont {P.}~\bibnamefont {Vieira}},\
  }\href@noop {} {\bibinfo {title} {{Structure Constants and Integrable
  Bootstrap in Planar N=4 SYM Theory}}} (\bibinfo {year} {2015}),\ \Eprint
  {https://arxiv.org/abs/1505.06745} {arXiv:1505.06745 [hep-th]} \BibitemShut
  {NoStop}%
\bibitem [{\citenamefont {Basso}\ \emph {et~al.}(2016)\citenamefont {Basso},
  \citenamefont {Goncalves}, \citenamefont {Komatsu},\ and\ \citenamefont
  {Vieira}}]{Basso:2015eqa}%
  \BibitemOpen
  \bibfield  {author} {\bibinfo {author} {\bibfnamefont {B.}~\bibnamefont
  {Basso}}, \bibinfo {author} {\bibfnamefont {V.}~\bibnamefont {Goncalves}},
  \bibinfo {author} {\bibfnamefont {S.}~\bibnamefont {Komatsu}},\ and\ \bibinfo
  {author} {\bibfnamefont {P.}~\bibnamefont {Vieira}},\ }\href
  {https://doi.org/10.1016/j.nuclphysb.2016.04.020} {\bibfield  {journal}
  {\bibinfo  {journal} {Nucl. Phys. B}\ }\textbf {\bibinfo {volume} {907}},\
  \bibinfo {pages} {695} (\bibinfo {year} {2016})},\ \Eprint
  {https://arxiv.org/abs/1510.01683} {arXiv:1510.01683 [hep-th]} \BibitemShut
  {NoStop}%
\bibitem [{\citenamefont {Eden}\ and\ \citenamefont
  {Sfondrini}(2016)}]{Eden:2015ija}%
  \BibitemOpen
  \bibfield  {author} {\bibinfo {author} {\bibfnamefont {B.}~\bibnamefont
  {Eden}}\ and\ \bibinfo {author} {\bibfnamefont {A.}~\bibnamefont
  {Sfondrini}},\ }\href {https://doi.org/10.1007/JHEP02(2016)165} {\bibfield
  {journal} {\bibinfo  {journal} {JHEP}\ }\textbf {\bibinfo {volume} {02}},\
  \bibinfo {pages} {165}},\ \Eprint {https://arxiv.org/abs/1510.01242}
  {arXiv:1510.01242 [hep-th]} \BibitemShut {NoStop}%
\bibitem [{\citenamefont {Basso}\ \emph {et~al.}(2017)\citenamefont {Basso},
  \citenamefont {Goncalves},\ and\ \citenamefont {Komatsu}}]{Basso:2017muf}%
  \BibitemOpen
  \bibfield  {author} {\bibinfo {author} {\bibfnamefont {B.}~\bibnamefont
  {Basso}}, \bibinfo {author} {\bibfnamefont {V.}~\bibnamefont {Goncalves}},\
  and\ \bibinfo {author} {\bibfnamefont {S.}~\bibnamefont {Komatsu}},\ }\href
  {https://doi.org/10.1007/JHEP05(2017)124} {\bibfield  {journal} {\bibinfo
  {journal} {JHEP}\ }\textbf {\bibinfo {volume} {05}},\ \bibinfo {pages}
  {124}},\ \Eprint {https://arxiv.org/abs/1702.02154} {arXiv:1702.02154
  [hep-th]} \BibitemShut {NoStop}%
\bibitem [{\citenamefont {Basso}\ \emph {et~al.}(2023)\citenamefont {Basso},
  \citenamefont {Georgoudis},\ and\ \citenamefont {Sueiro}}]{Basso:2022nny}%
  \BibitemOpen
  \bibfield  {author} {\bibinfo {author} {\bibfnamefont {B.}~\bibnamefont
  {Basso}}, \bibinfo {author} {\bibfnamefont {A.}~\bibnamefont {Georgoudis}},\
  and\ \bibinfo {author} {\bibfnamefont {A.~K.}\ \bibnamefont {Sueiro}},\
  }\href {https://doi.org/10.1103/PhysRevLett.130.131603} {\bibfield  {journal}
  {\bibinfo  {journal} {Phys. Rev. Lett.}\ }\textbf {\bibinfo {volume} {130}},\
  \bibinfo {pages} {131603} (\bibinfo {year} {2023})},\ \Eprint
  {https://arxiv.org/abs/2207.01315} {arXiv:2207.01315 [hep-th]} \BibitemShut
  {NoStop}%
\bibitem [{\citenamefont {Gromov}\ \emph {et~al.}(2016)\citenamefont {Gromov},
  \citenamefont {Levkovich-Maslyuk},\ and\ \citenamefont
  {Sizov}}]{Gromov:2015wca}%
  \BibitemOpen
  \bibfield  {author} {\bibinfo {author} {\bibfnamefont {N.}~\bibnamefont
  {Gromov}}, \bibinfo {author} {\bibfnamefont {F.}~\bibnamefont
  {Levkovich-Maslyuk}},\ and\ \bibinfo {author} {\bibfnamefont
  {G.}~\bibnamefont {Sizov}},\ }\href {https://doi.org/10.1007/JHEP06(2016)036}
  {\bibfield  {journal} {\bibinfo  {journal} {JHEP}\ }\textbf {\bibinfo
  {volume} {06}},\ \bibinfo {pages} {036}},\ \Eprint
  {https://arxiv.org/abs/1504.06640} {arXiv:1504.06640 [hep-th]} \BibitemShut
  {NoStop}%
\bibitem [{\citenamefont {Gromov}\ \emph {et~al.}(2024)\citenamefont {Gromov},
  \citenamefont {Hegedus}, \citenamefont {Julius},\ and\ \citenamefont
  {Sokolova}}]{Gromov:2023hzc}%
  \BibitemOpen
  \bibfield  {author} {\bibinfo {author} {\bibfnamefont {N.}~\bibnamefont
  {Gromov}}, \bibinfo {author} {\bibfnamefont {A.}~\bibnamefont {Hegedus}},
  \bibinfo {author} {\bibfnamefont {J.}~\bibnamefont {Julius}},\ and\ \bibinfo
  {author} {\bibfnamefont {N.}~\bibnamefont {Sokolova}},\ }\href
  {https://doi.org/10.1007/JHEP05(2024)185} {\bibfield  {journal} {\bibinfo
  {journal} {JHEP}\ }\textbf {\bibinfo {volume} {05}},\ \bibinfo {pages}
  {185}},\ \Eprint {https://arxiv.org/abs/2306.12379} {arXiv:2306.12379
  [hep-th]} \BibitemShut {NoStop}%
\bibitem [{\citenamefont {Ekhammar}\ \emph {et~al.}(2024)\citenamefont
  {Ekhammar}, \citenamefont {Gromov},\ and\ \citenamefont
  {Ryan}}]{Ekhammar:2024rfj}%
  \BibitemOpen
  \bibfield  {author} {\bibinfo {author} {\bibfnamefont {S.}~\bibnamefont
  {Ekhammar}}, \bibinfo {author} {\bibfnamefont {N.}~\bibnamefont {Gromov}},\
  and\ \bibinfo {author} {\bibfnamefont {P.}~\bibnamefont {Ryan}},\ }\href
  {https://doi.org/10.1007/JHEP12(2024)165} {\bibfield  {journal} {\bibinfo
  {journal} {JHEP}\ }\textbf {\bibinfo {volume} {12}},\ \bibinfo {pages}
  {165}},\ \Eprint {https://arxiv.org/abs/2406.02698} {arXiv:2406.02698
  [hep-th]} \BibitemShut {NoStop}%
\bibitem [{\citenamefont {Bornemann}(2010)}]{Bornemann:2010}%
  \BibitemOpen
  \bibfield  {author} {\bibinfo {author} {\bibfnamefont {F.}~\bibnamefont
  {Bornemann}},\ }\href {https://doi.org/10.1090/S0025-5718-09-02280-7}
  {\bibfield  {journal} {\bibinfo  {journal} {Math. Comput.}\ }\textbf
  {\bibinfo {volume} {79}},\ \bibinfo {pages} {871} (\bibinfo {year} {2010})},\
  \Eprint {https://arxiv.org/abs/0804.2543} {arXiv:0804.2543 [math.NA]}
  \BibitemShut {NoStop}%
\bibitem [{\citenamefont {Eden}(2012)}]{Eden:2012rr}%
  \BibitemOpen
  \bibfield  {author} {\bibinfo {author} {\bibfnamefont {B.}~\bibnamefont
  {Eden}},\ }\href@noop {} {\bibinfo {title} {{Three-loop universal structure
  constants in N=4 susy Yang-Mills theory}}} (\bibinfo {year} {2012}),\ \Eprint
  {https://arxiv.org/abs/1207.3112} {arXiv:1207.3112 [hep-th]} \BibitemShut
  {NoStop}%
\bibitem [{\citenamefont {Eden}\ and\ \citenamefont
  {Paul}(2016)}]{Eden:2016aqo}%
  \BibitemOpen
  \bibfield  {author} {\bibinfo {author} {\bibfnamefont {B.}~\bibnamefont
  {Eden}}\ and\ \bibinfo {author} {\bibfnamefont {F.}~\bibnamefont {Paul}},\
  }\href@noop {} {\bibinfo {title} {{Half-BPS half-BPS twist two at four loops
  in N=4 SYM}}} (\bibinfo {year} {2016}),\ \Eprint
  {https://arxiv.org/abs/1608.04222} {arXiv:1608.04222 [hep-th]} \BibitemShut
  {NoStop}%
\bibitem [{\citenamefont {Gon{\c{c}}alves}(2017)}]{Goncalves:2016vir}%
  \BibitemOpen
  \bibfield  {author} {\bibinfo {author} {\bibfnamefont {V.}~\bibnamefont
  {Gon{\c{c}}alves}},\ }\href {https://doi.org/10.1007/JHEP03(2017)079}
  {\bibfield  {journal} {\bibinfo  {journal} {JHEP}\ }\textbf {\bibinfo
  {volume} {03}},\ \bibinfo {pages} {079}},\ \Eprint
  {https://arxiv.org/abs/1607.02195} {arXiv:1607.02195 [hep-th]} \BibitemShut
  {NoStop}%
\bibitem [{\citenamefont {Georgoudis}\ \emph {et~al.}(2018)\citenamefont
  {Georgoudis}, \citenamefont {Goncalves},\ and\ \citenamefont
  {Pereira}}]{Georgoudis:2017meq}%
  \BibitemOpen
  \bibfield  {author} {\bibinfo {author} {\bibfnamefont {A.}~\bibnamefont
  {Georgoudis}}, \bibinfo {author} {\bibfnamefont {V.}~\bibnamefont
  {Goncalves}},\ and\ \bibinfo {author} {\bibfnamefont {R.}~\bibnamefont
  {Pereira}},\ }\href {https://doi.org/10.1007/JHEP11(2018)184} {\bibfield
  {journal} {\bibinfo  {journal} {JHEP}\ }\textbf {\bibinfo {volume} {11}},\
  \bibinfo {pages} {184}},\ \Eprint {https://arxiv.org/abs/1710.06419}
  {arXiv:1710.06419 [hep-th]} \BibitemShut {NoStop}%
\bibitem [{\citenamefont {Chicherin}\ \emph {et~al.}(2018)\citenamefont
  {Chicherin}, \citenamefont {Georgoudis}, \citenamefont {Gon{\c{c}}alves},\
  and\ \citenamefont {Pereira}}]{Chicherin:2018avq}%
  \BibitemOpen
  \bibfield  {author} {\bibinfo {author} {\bibfnamefont {D.}~\bibnamefont
  {Chicherin}}, \bibinfo {author} {\bibfnamefont {A.}~\bibnamefont
  {Georgoudis}}, \bibinfo {author} {\bibfnamefont {V.}~\bibnamefont
  {Gon{\c{c}}alves}},\ and\ \bibinfo {author} {\bibfnamefont {R.}~\bibnamefont
  {Pereira}},\ }\href {https://doi.org/10.1007/JHEP11(2018)069} {\bibfield
  {journal} {\bibinfo  {journal} {JHEP}\ }\textbf {\bibinfo {volume} {11}},\
  \bibinfo {pages} {069}},\ \Eprint {https://arxiv.org/abs/1809.00551}
  {arXiv:1809.00551 [hep-th]} \BibitemShut {NoStop}%
\bibitem [{\citenamefont {Bargheer}\ \emph {et~al.}(2014)\citenamefont
  {Bargheer}, \citenamefont {Minahan},\ and\ \citenamefont
  {Pereira}}]{Bargheer:2013faa}%
  \BibitemOpen
  \bibfield  {author} {\bibinfo {author} {\bibfnamefont {T.}~\bibnamefont
  {Bargheer}}, \bibinfo {author} {\bibfnamefont {J.~A.}\ \bibnamefont
  {Minahan}},\ and\ \bibinfo {author} {\bibfnamefont {R.}~\bibnamefont
  {Pereira}},\ }\href {https://doi.org/10.1007/JHEP03(2014)096} {\bibfield
  {journal} {\bibinfo  {journal} {JHEP}\ }\textbf {\bibinfo {volume} {03}},\
  \bibinfo {pages} {096}},\ \Eprint {https://arxiv.org/abs/1311.7461}
  {arXiv:1311.7461 [hep-th]} \BibitemShut {NoStop}%
\bibitem [{\citenamefont {Gon{\c{c}}alves}(2015)}]{Goncalves:2014ffa}%
  \BibitemOpen
  \bibfield  {author} {\bibinfo {author} {\bibfnamefont {V.}~\bibnamefont
  {Gon{\c{c}}alves}},\ }\href {https://doi.org/10.1007/JHEP04(2015)150}
  {\bibfield  {journal} {\bibinfo  {journal} {JHEP}\ }\textbf {\bibinfo
  {volume} {04}},\ \bibinfo {pages} {150}},\ \Eprint
  {https://arxiv.org/abs/1411.1675} {arXiv:1411.1675 [hep-th]} \BibitemShut
  {NoStop}%
\bibitem [{\citenamefont {Alday}\ \emph {et~al.}(2023)\citenamefont {Alday},
  \citenamefont {Hansen},\ and\ \citenamefont {Silva}}]{Alday:2023flc}%
  \BibitemOpen
  \bibfield  {author} {\bibinfo {author} {\bibfnamefont {L.~F.}\ \bibnamefont
  {Alday}}, \bibinfo {author} {\bibfnamefont {T.}~\bibnamefont {Hansen}},\ and\
  \bibinfo {author} {\bibfnamefont {J.~A.}\ \bibnamefont {Silva}},\ }\href
  {https://doi.org/10.1007/JHEP08(2023)214} {\bibfield  {journal} {\bibinfo
  {journal} {JHEP}\ }\textbf {\bibinfo {volume} {08}},\ \bibinfo {pages}
  {214}},\ \Eprint {https://arxiv.org/abs/2303.08834} {arXiv:2303.08834
  [hep-th]} \BibitemShut {NoStop}%
\bibitem [{\citenamefont {Beem}\ \emph {et~al.}(2013)\citenamefont {Beem},
  \citenamefont {Rastelli},\ and\ \citenamefont {van Rees}}]{Beem:2013qxa}%
  \BibitemOpen
  \bibfield  {author} {\bibinfo {author} {\bibfnamefont {C.}~\bibnamefont
  {Beem}}, \bibinfo {author} {\bibfnamefont {L.}~\bibnamefont {Rastelli}},\
  and\ \bibinfo {author} {\bibfnamefont {B.~C.}\ \bibnamefont {van Rees}},\
  }\href {https://doi.org/10.1103/PhysRevLett.111.071601} {\bibfield  {journal}
  {\bibinfo  {journal} {Phys. Rev. Lett.}\ }\textbf {\bibinfo {volume} {111}},\
  \bibinfo {pages} {071601} (\bibinfo {year} {2013})},\ \Eprint
  {https://arxiv.org/abs/1304.1803} {arXiv:1304.1803 [hep-th]} \BibitemShut
  {NoStop}%
\bibitem [{\citenamefont {Beem}\ \emph {et~al.}(2017)\citenamefont {Beem},
  \citenamefont {Rastelli},\ and\ \citenamefont {van Rees}}]{Beem:2016wfs}%
  \BibitemOpen
  \bibfield  {author} {\bibinfo {author} {\bibfnamefont {C.}~\bibnamefont
  {Beem}}, \bibinfo {author} {\bibfnamefont {L.}~\bibnamefont {Rastelli}},\
  and\ \bibinfo {author} {\bibfnamefont {B.~C.}\ \bibnamefont {van Rees}},\
  }\href {https://doi.org/10.1103/PhysRevD.96.046014} {\bibfield  {journal}
  {\bibinfo  {journal} {Phys. Rev. D}\ }\textbf {\bibinfo {volume} {96}},\
  \bibinfo {pages} {046014} (\bibinfo {year} {2017})},\ \Eprint
  {https://arxiv.org/abs/1612.02363} {arXiv:1612.02363 [hep-th]} \BibitemShut
  {NoStop}%
\bibitem [{\citenamefont {Caron-Huot}\ \emph {et~al.}(2023)\citenamefont
  {Caron-Huot}, \citenamefont {Coronado}, \citenamefont {Trinh},\ and\
  \citenamefont {Zahraee}}]{Caron-Huot:2022sdy}%
  \BibitemOpen
  \bibfield  {author} {\bibinfo {author} {\bibfnamefont {S.}~\bibnamefont
  {Caron-Huot}}, \bibinfo {author} {\bibfnamefont {F.}~\bibnamefont
  {Coronado}}, \bibinfo {author} {\bibfnamefont {A.-K.}\ \bibnamefont
  {Trinh}},\ and\ \bibinfo {author} {\bibfnamefont {Z.}~\bibnamefont
  {Zahraee}},\ }\href {https://doi.org/10.1007/JHEP02(2023)083} {\bibfield
  {journal} {\bibinfo  {journal} {JHEP}\ }\textbf {\bibinfo {volume} {02}},\
  \bibinfo {pages} {083}},\ \Eprint {https://arxiv.org/abs/2207.01615}
  {arXiv:2207.01615 [hep-th]} \BibitemShut {NoStop}%
\bibitem [{\citenamefont {Minahan}\ and\ \citenamefont
  {Zarembo}(2003)}]{Minahan:2002ve}%
  \BibitemOpen
  \bibfield  {author} {\bibinfo {author} {\bibfnamefont {J.~A.}\ \bibnamefont
  {Minahan}}\ and\ \bibinfo {author} {\bibfnamefont {K.}~\bibnamefont
  {Zarembo}},\ }\href {https://doi.org/10.1088/1126-6708/2003/03/013}
  {\bibfield  {journal} {\bibinfo  {journal} {JHEP}\ }\textbf {\bibinfo
  {volume} {03}},\ \bibinfo {pages} {013}},\ \Eprint
  {https://arxiv.org/abs/hep-th/0212208} {arXiv:hep-th/0212208} \BibitemShut
  {NoStop}%
\bibitem [{\citenamefont {Beisert}\ and\ \citenamefont
  {Staudacher}(2003)}]{Beisert:2003yb}%
  \BibitemOpen
  \bibfield  {author} {\bibinfo {author} {\bibfnamefont {N.}~\bibnamefont
  {Beisert}}\ and\ \bibinfo {author} {\bibfnamefont {M.}~\bibnamefont
  {Staudacher}},\ }\href {https://doi.org/10.1016/j.nuclphysb.2003.08.015}
  {\bibfield  {journal} {\bibinfo  {journal} {Nucl. Phys. B}\ }\textbf
  {\bibinfo {volume} {670}},\ \bibinfo {pages} {439} (\bibinfo {year}
  {2003})},\ \Eprint {https://arxiv.org/abs/hep-th/0307042}
  {arXiv:hep-th/0307042} \BibitemShut {NoStop}%
\bibitem [{\citenamefont {Beisert}(2008)}]{Beisert:2005tm}%
  \BibitemOpen
  \bibfield  {author} {\bibinfo {author} {\bibfnamefont {N.}~\bibnamefont
  {Beisert}},\ }\href {https://doi.org/10.4310/ATMP.2008.v12.n5.a1} {\bibfield
  {journal} {\bibinfo  {journal} {Adv. Theor. Math. Phys.}\ }\textbf {\bibinfo
  {volume} {12}},\ \bibinfo {pages} {945} (\bibinfo {year} {2008})},\ \Eprint
  {https://arxiv.org/abs/hep-th/0511082} {arXiv:hep-th/0511082} \BibitemShut
  {NoStop}%
\bibitem [{\citenamefont {Janik}(2006)}]{Janik:2006dc}%
  \BibitemOpen
  \bibfield  {author} {\bibinfo {author} {\bibfnamefont {R.~A.}\ \bibnamefont
  {Janik}},\ }\href {https://doi.org/10.1103/PhysRevD.73.086006} {\bibfield
  {journal} {\bibinfo  {journal} {Phys. Rev. D}\ }\textbf {\bibinfo {volume}
  {73}},\ \bibinfo {pages} {086006} (\bibinfo {year} {2006})},\ \Eprint
  {https://arxiv.org/abs/hep-th/0603038} {arXiv:hep-th/0603038} \BibitemShut
  {NoStop}%
\bibitem [{\citenamefont {Beisert}\ and\ \citenamefont
  {Staudacher}(2005)}]{Beisert:2005fw}%
  \BibitemOpen
  \bibfield  {author} {\bibinfo {author} {\bibfnamefont {N.}~\bibnamefont
  {Beisert}}\ and\ \bibinfo {author} {\bibfnamefont {M.}~\bibnamefont
  {Staudacher}},\ }\href {https://doi.org/10.1016/j.nuclphysb.2005.06.038}
  {\bibfield  {journal} {\bibinfo  {journal} {Nucl. Phys. B}\ }\textbf
  {\bibinfo {volume} {727}},\ \bibinfo {pages} {1} (\bibinfo {year} {2005})},\
  \Eprint {https://arxiv.org/abs/hep-th/0504190} {arXiv:hep-th/0504190}
  \BibitemShut {NoStop}%
\bibitem [{\citenamefont {Beisert}\ \emph {et~al.}(2007)\citenamefont
  {Beisert}, \citenamefont {Eden},\ and\ \citenamefont
  {Staudacher}}]{Beisert:2006ez}%
  \BibitemOpen
  \bibfield  {author} {\bibinfo {author} {\bibfnamefont {N.}~\bibnamefont
  {Beisert}}, \bibinfo {author} {\bibfnamefont {B.}~\bibnamefont {Eden}},\ and\
  \bibinfo {author} {\bibfnamefont {M.}~\bibnamefont {Staudacher}},\ }\href
  {https://doi.org/10.1088/1742-5468/2007/01/P01021} {\bibfield  {journal}
  {\bibinfo  {journal} {J. Stat. Mech.}\ }\textbf {\bibinfo {volume} {0701}},\
  \bibinfo {pages} {P01021} (\bibinfo {year} {2007})},\ \Eprint
  {https://arxiv.org/abs/hep-th/0610251} {arXiv:hep-th/0610251} \BibitemShut
  {NoStop}%
\bibitem [{\citenamefont {Kotikov}\ \emph {et~al.}(2007)\citenamefont
  {Kotikov}, \citenamefont {Lipatov}, \citenamefont {Rej}, \citenamefont
  {Staudacher},\ and\ \citenamefont {Velizhanin}}]{Kotikov:2007cy}%
  \BibitemOpen
  \bibfield  {author} {\bibinfo {author} {\bibfnamefont {A.~V.}\ \bibnamefont
  {Kotikov}}, \bibinfo {author} {\bibfnamefont {L.~N.}\ \bibnamefont
  {Lipatov}}, \bibinfo {author} {\bibfnamefont {A.}~\bibnamefont {Rej}},
  \bibinfo {author} {\bibfnamefont {M.}~\bibnamefont {Staudacher}},\ and\
  \bibinfo {author} {\bibfnamefont {V.~N.}\ \bibnamefont {Velizhanin}},\ }\href
  {https://doi.org/10.1088/1742-5468/2007/10/P10003} {\bibfield  {journal}
  {\bibinfo  {journal} {J. Stat. Mech.}\ }\textbf {\bibinfo {volume} {0710}},\
  \bibinfo {pages} {P10003} (\bibinfo {year} {2007})},\ \Eprint
  {https://arxiv.org/abs/0704.3586} {arXiv:0704.3586 [hep-th]} \BibitemShut
  {NoStop}%
\bibitem [{\citenamefont {Bajnok}\ and\ \citenamefont
  {Janik}(2009)}]{Bajnok:2008bm}%
  \BibitemOpen
  \bibfield  {author} {\bibinfo {author} {\bibfnamefont {Z.}~\bibnamefont
  {Bajnok}}\ and\ \bibinfo {author} {\bibfnamefont {R.~A.}\ \bibnamefont
  {Janik}},\ }\href {https://doi.org/10.1016/j.nuclphysb.2008.08.020}
  {\bibfield  {journal} {\bibinfo  {journal} {Nucl. Phys. B}\ }\textbf
  {\bibinfo {volume} {807}},\ \bibinfo {pages} {625} (\bibinfo {year}
  {2009})},\ \Eprint {https://arxiv.org/abs/0807.0399} {arXiv:0807.0399
  [hep-th]} \BibitemShut {NoStop}%
\bibitem [{\citenamefont {Bombardelli}\ \emph {et~al.}(2009)\citenamefont
  {Bombardelli}, \citenamefont {Fioravanti},\ and\ \citenamefont
  {Tateo}}]{Bombardelli:2009ns}%
  \BibitemOpen
  \bibfield  {author} {\bibinfo {author} {\bibfnamefont {D.}~\bibnamefont
  {Bombardelli}}, \bibinfo {author} {\bibfnamefont {D.}~\bibnamefont
  {Fioravanti}},\ and\ \bibinfo {author} {\bibfnamefont {R.}~\bibnamefont
  {Tateo}},\ }\href {https://doi.org/10.1088/1751-8113/42/37/375401} {\bibfield
   {journal} {\bibinfo  {journal} {J. Phys. A}\ }\textbf {\bibinfo {volume}
  {42}},\ \bibinfo {pages} {375401} (\bibinfo {year} {2009})},\ \Eprint
  {https://arxiv.org/abs/0902.3930} {arXiv:0902.3930 [hep-th]} \BibitemShut
  {NoStop}%
\bibitem [{\citenamefont {Gromov}\ \emph {et~al.}(2009)\citenamefont {Gromov},
  \citenamefont {Kazakov},\ and\ \citenamefont {Vieira}}]{Gromov:2009tv}%
  \BibitemOpen
  \bibfield  {author} {\bibinfo {author} {\bibfnamefont {N.}~\bibnamefont
  {Gromov}}, \bibinfo {author} {\bibfnamefont {V.}~\bibnamefont {Kazakov}},\
  and\ \bibinfo {author} {\bibfnamefont {P.}~\bibnamefont {Vieira}},\ }\href
  {https://doi.org/10.1103/PhysRevLett.103.131601} {\bibfield  {journal}
  {\bibinfo  {journal} {Phys. Rev. Lett.}\ }\textbf {\bibinfo {volume} {103}},\
  \bibinfo {pages} {131601} (\bibinfo {year} {2009})},\ \Eprint
  {https://arxiv.org/abs/0901.3753} {arXiv:0901.3753 [hep-th]} \BibitemShut
  {NoStop}%
\bibitem [{\citenamefont {Arutyunov}\ and\ \citenamefont
  {Frolov}(2009)}]{Arutyunov:2009ur}%
  \BibitemOpen
  \bibfield  {author} {\bibinfo {author} {\bibfnamefont {G.}~\bibnamefont
  {Arutyunov}}\ and\ \bibinfo {author} {\bibfnamefont {S.}~\bibnamefont
  {Frolov}},\ }\href {https://doi.org/10.1088/1126-6708/2009/05/068} {\bibfield
   {journal} {\bibinfo  {journal} {JHEP}\ }\textbf {\bibinfo {volume} {05}},\
  \bibinfo {pages} {068}},\ \Eprint {https://arxiv.org/abs/0903.0141}
  {arXiv:0903.0141 [hep-th]} \BibitemShut {NoStop}%
\bibitem [{\citenamefont {Marboe}\ and\ \citenamefont
  {Volin}(2015)}]{Marboe:2014gma}%
  \BibitemOpen
  \bibfield  {author} {\bibinfo {author} {\bibfnamefont {C.}~\bibnamefont
  {Marboe}}\ and\ \bibinfo {author} {\bibfnamefont {D.}~\bibnamefont {Volin}},\
  }\href {https://doi.org/10.1016/j.nuclphysb.2015.08.021} {\bibfield
  {journal} {\bibinfo  {journal} {Nucl. Phys. B}\ }\textbf {\bibinfo {volume}
  {899}},\ \bibinfo {pages} {810} (\bibinfo {year} {2015})},\ \Eprint
  {https://arxiv.org/abs/1411.4758} {arXiv:1411.4758 [hep-th]} \BibitemShut
  {NoStop}%
\bibitem [{\citenamefont {Marboe}\ and\ \citenamefont
  {Volin}(2021)}]{Marboe:2018ugv}%
  \BibitemOpen
  \bibfield  {author} {\bibinfo {author} {\bibfnamefont {C.}~\bibnamefont
  {Marboe}}\ and\ \bibinfo {author} {\bibfnamefont {D.}~\bibnamefont {Volin}},\
  }\href {https://doi.org/10.1088/1751-8121/abd59c} {\bibfield  {journal}
  {\bibinfo  {journal} {J. Phys. A}\ }\textbf {\bibinfo {volume} {54}},\
  \bibinfo {pages} {055201} (\bibinfo {year} {2021})},\ \Eprint
  {https://arxiv.org/abs/1812.09238} {arXiv:1812.09238 [hep-th]} \BibitemShut
  {NoStop}%
\bibitem [{\citenamefont {Gromov}\ \emph
  {et~al.}(2014{\natexlab{b}})\citenamefont {Gromov}, \citenamefont
  {Levkovich-Maslyuk}, \citenamefont {Sizov},\ and\ \citenamefont
  {Valatka}}]{Gromov:2014bva}%
  \BibitemOpen
  \bibfield  {author} {\bibinfo {author} {\bibfnamefont {N.}~\bibnamefont
  {Gromov}}, \bibinfo {author} {\bibfnamefont {F.}~\bibnamefont
  {Levkovich-Maslyuk}}, \bibinfo {author} {\bibfnamefont {G.}~\bibnamefont
  {Sizov}},\ and\ \bibinfo {author} {\bibfnamefont {S.}~\bibnamefont
  {Valatka}},\ }\href {https://doi.org/10.1007/JHEP07(2014)156} {\bibfield
  {journal} {\bibinfo  {journal} {JHEP}\ }\textbf {\bibinfo {volume} {07}},\
  \bibinfo {pages} {156}},\ \Eprint {https://arxiv.org/abs/1402.0871}
  {arXiv:1402.0871 [hep-th]} \BibitemShut {NoStop}%
\bibitem [{\citenamefont {Sklyanin}(1995)}]{Sklyanin:1995bm}%
  \BibitemOpen
  \bibfield  {author} {\bibinfo {author} {\bibfnamefont {E.~K.}\ \bibnamefont
  {Sklyanin}},\ }\href {https://doi.org/10.1143/PTPS.118.35} {\bibfield
  {journal} {\bibinfo  {journal} {Prog. Theor. Phys. Suppl.}\ }\textbf
  {\bibinfo {volume} {118}},\ \bibinfo {pages} {35} (\bibinfo {year} {1995})},\
  \Eprint {https://arxiv.org/abs/solv-int/9504001} {arXiv:solv-int/9504001}
  \BibitemShut {NoStop}%
\bibitem [{\citenamefont {Gromov}\ \emph {et~al.}(2017)\citenamefont {Gromov},
  \citenamefont {Levkovich-Maslyuk},\ and\ \citenamefont
  {Sizov}}]{Gromov:2016itr}%
  \BibitemOpen
  \bibfield  {author} {\bibinfo {author} {\bibfnamefont {N.}~\bibnamefont
  {Gromov}}, \bibinfo {author} {\bibfnamefont {F.}~\bibnamefont
  {Levkovich-Maslyuk}},\ and\ \bibinfo {author} {\bibfnamefont
  {G.}~\bibnamefont {Sizov}},\ }\href {https://doi.org/10.1007/JHEP09(2017)111}
  {\bibfield  {journal} {\bibinfo  {journal} {JHEP}\ }\textbf {\bibinfo
  {volume} {09}},\ \bibinfo {pages} {111}},\ \Eprint
  {https://arxiv.org/abs/1610.08032} {arXiv:1610.08032 [hep-th]} \BibitemShut
  {NoStop}%
\bibitem [{\citenamefont {Ryan}\ and\ \citenamefont
  {Volin}(2019)}]{Ryan:2018fyo}%
  \BibitemOpen
  \bibfield  {author} {\bibinfo {author} {\bibfnamefont {P.}~\bibnamefont
  {Ryan}}\ and\ \bibinfo {author} {\bibfnamefont {D.}~\bibnamefont {Volin}},\
  }\href {https://doi.org/10.1063/1.5085387} {\bibfield  {journal} {\bibinfo
  {journal} {J. Math. Phys.}\ }\textbf {\bibinfo {volume} {60}},\ \bibinfo
  {pages} {032701} (\bibinfo {year} {2019})},\ \Eprint
  {https://arxiv.org/abs/1810.10996} {arXiv:1810.10996 [math-ph]} \BibitemShut
  {NoStop}%
\bibitem [{\citenamefont {Cavagli{\`a}}\ \emph {et~al.}(2019)\citenamefont
  {Cavagli{\`a}}, \citenamefont {Gromov},\ and\ \citenamefont
  {Levkovich-Maslyuk}}]{Cavaglia:2019pow}%
  \BibitemOpen
  \bibfield  {author} {\bibinfo {author} {\bibfnamefont {A.}~\bibnamefont
  {Cavagli{\`a}}}, \bibinfo {author} {\bibfnamefont {N.}~\bibnamefont
  {Gromov}},\ and\ \bibinfo {author} {\bibfnamefont {F.}~\bibnamefont
  {Levkovich-Maslyuk}},\ }\href {https://doi.org/10.1007/JHEP09(2019)052}
  {\bibfield  {journal} {\bibinfo  {journal} {JHEP}\ }\textbf {\bibinfo
  {volume} {09}},\ \bibinfo {pages} {052}},\ \Eprint
  {https://arxiv.org/abs/1907.03788} {arXiv:1907.03788 [hep-th]} \BibitemShut
  {NoStop}%
\bibitem [{\citenamefont {Gromov}\ \emph {et~al.}(2021)\citenamefont {Gromov},
  \citenamefont {Levkovich-Maslyuk},\ and\ \citenamefont
  {Ryan}}]{Gromov:2020fwh}%
  \BibitemOpen
  \bibfield  {author} {\bibinfo {author} {\bibfnamefont {N.}~\bibnamefont
  {Gromov}}, \bibinfo {author} {\bibfnamefont {F.}~\bibnamefont
  {Levkovich-Maslyuk}},\ and\ \bibinfo {author} {\bibfnamefont
  {P.}~\bibnamefont {Ryan}},\ }\href {https://doi.org/10.1007/JHEP05(2021)169}
  {\bibfield  {journal} {\bibinfo  {journal} {JHEP}\ }\textbf {\bibinfo
  {volume} {05}},\ \bibinfo {pages} {169}},\ \Eprint
  {https://arxiv.org/abs/2011.08229} {arXiv:2011.08229 [hep-th]} \BibitemShut
  {NoStop}%
\bibitem [{\citenamefont {Gromov}\ \emph {et~al.}(2022)\citenamefont {Gromov},
  \citenamefont {Primi},\ and\ \citenamefont {Ryan}}]{Gromov:2022waj}%
  \BibitemOpen
  \bibfield  {author} {\bibinfo {author} {\bibfnamefont {N.}~\bibnamefont
  {Gromov}}, \bibinfo {author} {\bibfnamefont {N.}~\bibnamefont {Primi}},\ and\
  \bibinfo {author} {\bibfnamefont {P.}~\bibnamefont {Ryan}},\ }\href
  {https://doi.org/10.1007/JHEP11(2022)039} {\bibfield  {journal} {\bibinfo
  {journal} {JHEP}\ }\textbf {\bibinfo {volume} {11}},\ \bibinfo {pages}
  {039}},\ \Eprint {https://arxiv.org/abs/2202.01591} {arXiv:2202.01591
  [hep-th]} \BibitemShut {NoStop}%
\bibitem [{\citenamefont {Cavagli{\`a}}\ \emph {et~al.}(2018)\citenamefont
  {Cavagli{\`a}}, \citenamefont {Gromov},\ and\ \citenamefont
  {Levkovich-Maslyuk}}]{Cavaglia:2018lxi}%
  \BibitemOpen
  \bibfield  {author} {\bibinfo {author} {\bibfnamefont {A.}~\bibnamefont
  {Cavagli{\`a}}}, \bibinfo {author} {\bibfnamefont {N.}~\bibnamefont
  {Gromov}},\ and\ \bibinfo {author} {\bibfnamefont {F.}~\bibnamefont
  {Levkovich-Maslyuk}},\ }\href {https://doi.org/10.1007/JHEP10(2018)060}
  {\bibfield  {journal} {\bibinfo  {journal} {JHEP}\ }\textbf {\bibinfo
  {volume} {10}},\ \bibinfo {pages} {060}},\ \Eprint
  {https://arxiv.org/abs/1802.04237} {arXiv:1802.04237 [hep-th]} \BibitemShut
  {NoStop}%
\bibitem [{\citenamefont {Giombi}\ and\ \citenamefont
  {Komatsu}(2018)}]{Giombi:2018qox}%
  \BibitemOpen
  \bibfield  {author} {\bibinfo {author} {\bibfnamefont {S.}~\bibnamefont
  {Giombi}}\ and\ \bibinfo {author} {\bibfnamefont {S.}~\bibnamefont
  {Komatsu}},\ }\href {https://doi.org/10.1007/JHEP05(2018)109} {\bibfield
  {journal} {\bibinfo  {journal} {JHEP}\ }\textbf {\bibinfo {volume} {05}},\
  \bibinfo {pages} {109}},\ \bibinfo {note} {[Erratum: JHEP 11, 123 (2018)]},\
  \Eprint {https://arxiv.org/abs/1802.05201} {arXiv:1802.05201 [hep-th]}
  \BibitemShut {NoStop}%
\bibitem [{\citenamefont {Cavagli{\`a}}\ \emph
  {et~al.}(2021{\natexlab{a}})\citenamefont {Cavagli{\`a}}, \citenamefont
  {Gromov},\ and\ \citenamefont {Levkovich-Maslyuk}}]{Cavaglia:2021mft}%
  \BibitemOpen
  \bibfield  {author} {\bibinfo {author} {\bibfnamefont {A.}~\bibnamefont
  {Cavagli{\`a}}}, \bibinfo {author} {\bibfnamefont {N.}~\bibnamefont
  {Gromov}},\ and\ \bibinfo {author} {\bibfnamefont {F.}~\bibnamefont
  {Levkovich-Maslyuk}},\ }\href {https://doi.org/10.1007/JHEP06(2021)131}
  {\bibfield  {journal} {\bibinfo  {journal} {JHEP}\ }\textbf {\bibinfo
  {volume} {06}},\ \bibinfo {pages} {131}},\ \Eprint
  {https://arxiv.org/abs/2103.15800} {arXiv:2103.15800 [hep-th]} \BibitemShut
  {NoStop}%
\bibitem [{\citenamefont {Bercini}\ \emph {et~al.}(2026)\citenamefont
  {Bercini}, \citenamefont {Ekhammar}, \citenamefont {Gromov},\ and\
  \citenamefont {Ryan}}]{Upcoming2D}%
  \BibitemOpen
  \bibfield  {author} {\bibinfo {author} {\bibfnamefont {C.}~\bibnamefont
  {Bercini}}, \bibinfo {author} {\bibfnamefont {S.}~\bibnamefont {Ekhammar}},
  \bibinfo {author} {\bibfnamefont {N.}~\bibnamefont {Gromov}},\ and\ \bibinfo
  {author} {\bibfnamefont {P.}~\bibnamefont {Ryan}}} (\bibinfo {year} {2026}),\
  \bibinfo {note} {in preparation}\BibitemShut {NoStop}%
\bibitem [{\citenamefont {Bercini}\ \emph {et~al.}(2024)\citenamefont
  {Bercini}, \citenamefont {Homrich},\ and\ \citenamefont
  {Vieira}}]{Bercini:2022jxo}%
  \BibitemOpen
  \bibfield  {author} {\bibinfo {author} {\bibfnamefont {C.}~\bibnamefont
  {Bercini}}, \bibinfo {author} {\bibfnamefont {A.}~\bibnamefont {Homrich}},\
  and\ \bibinfo {author} {\bibfnamefont {P.}~\bibnamefont {Vieira}},\ }\href
  {https://doi.org/10.1103/PhysRevD.110.L121901} {\bibfield  {journal}
  {\bibinfo  {journal} {Phys. Rev. D}\ }\textbf {\bibinfo {volume} {110}},\
  \bibinfo {pages} {L121901} (\bibinfo {year} {2024})},\ \Eprint
  {https://arxiv.org/abs/2210.04923} {arXiv:2210.04923 [hep-th]} \BibitemShut
  {NoStop}%
\bibitem [{\citenamefont {Bargheer}\ \emph {et~al.}(2026)\citenamefont
  {Bargheer}, \citenamefont {Bercini}, \citenamefont {Lefundes},\ and\
  \citenamefont {Ryan}}]{Bargheer:2026kon}%
  \BibitemOpen
  \bibfield  {author} {\bibinfo {author} {\bibfnamefont {T.}~\bibnamefont
  {Bargheer}}, \bibinfo {author} {\bibfnamefont {C.}~\bibnamefont {Bercini}},
  \bibinfo {author} {\bibfnamefont {G.}~\bibnamefont {Lefundes}},\ and\
  \bibinfo {author} {\bibfnamefont {P.}~\bibnamefont {Ryan}},\ }\href
  {https://doi.org/10.1103/tm1j-wxq7} {\bibfield  {journal} {\bibinfo
  {journal} {Phys. Rev. D}\ }\textbf {\bibinfo {volume} {114}},\ \bibinfo
  {pages} {L081904} (\bibinfo {year} {2026})},\ \Eprint
  {https://arxiv.org/abs/2603.26617} {arXiv:2603.26617 [hep-th]} \BibitemShut
  {NoStop}%
\bibitem [{\citenamefont {Basso}\ and\ \citenamefont
  {Georgoudis}(2025)}]{Basso:2025mca}%
  \BibitemOpen
  \bibfield  {author} {\bibinfo {author} {\bibfnamefont {B.}~\bibnamefont
  {Basso}}\ and\ \bibinfo {author} {\bibfnamefont {A.}~\bibnamefont
  {Georgoudis}},\ }\href {https://doi.org/10.1007/JHEP12(2025)119} {\bibfield
  {journal} {\bibinfo  {journal} {JHEP}\ }\textbf {\bibinfo {volume} {12}},\
  \bibinfo {pages} {119}},\ \Eprint {https://arxiv.org/abs/2504.07922}
  {arXiv:2504.07922 [hep-th]} \BibitemShut {NoStop}%
\bibitem [{\citenamefont {Jiang}\ \emph {et~al.}(2016)\citenamefont {Jiang},
  \citenamefont {Komatsu}, \citenamefont {Kostov},\ and\ \citenamefont
  {Serban}}]{Jiang:2016ulr}%
  \BibitemOpen
  \bibfield  {author} {\bibinfo {author} {\bibfnamefont {Y.}~\bibnamefont
  {Jiang}}, \bibinfo {author} {\bibfnamefont {S.}~\bibnamefont {Komatsu}},
  \bibinfo {author} {\bibfnamefont {I.}~\bibnamefont {Kostov}},\ and\ \bibinfo
  {author} {\bibfnamefont {D.}~\bibnamefont {Serban}},\ }\href
  {https://doi.org/10.1088/1751-8113/49/45/454003} {\bibfield  {journal}
  {\bibinfo  {journal} {J. Phys. A}\ }\textbf {\bibinfo {volume} {49}},\
  \bibinfo {pages} {454003} (\bibinfo {year} {2016})},\ \Eprint
  {https://arxiv.org/abs/1604.03575} {arXiv:1604.03575 [hep-th]} \BibitemShut
  {NoStop}%
\bibitem [{\citenamefont {Georgoudis}(2026)}]{upcomingLS}%
  \BibitemOpen
  \bibfield  {author} {\bibinfo {author} {\bibfnamefont {A.}~\bibnamefont
  {Georgoudis}}} (\bibinfo {year} {2026}),\ \bibinfo {note} {in
  preparation}\BibitemShut {NoStop}%
\bibitem [{\citenamefont {Basso}\ and\ \citenamefont
  {Zhong}(2020)}]{Basso:2019diw}%
  \BibitemOpen
  \bibfield  {author} {\bibinfo {author} {\bibfnamefont {B.}~\bibnamefont
  {Basso}}\ and\ \bibinfo {author} {\bibfnamefont {D.-L.}\ \bibnamefont
  {Zhong}},\ }\href {https://doi.org/10.1007/JHEP04(2020)076} {\bibfield
  {journal} {\bibinfo  {journal} {JHEP}\ }\textbf {\bibinfo {volume} {04}},\
  \bibinfo {pages} {076}},\ \Eprint {https://arxiv.org/abs/1907.01534}
  {arXiv:1907.01534 [hep-th]} \BibitemShut {NoStop}%
\bibitem [{\citenamefont {Gromov}\ and\ \citenamefont
  {Valatka}(2012)}]{Gromov:2011bz}%
  \BibitemOpen
  \bibfield  {author} {\bibinfo {author} {\bibfnamefont {N.}~\bibnamefont
  {Gromov}}\ and\ \bibinfo {author} {\bibfnamefont {S.}~\bibnamefont
  {Valatka}},\ }\href {https://doi.org/10.1007/JHEP03(2012)058} {\bibfield
  {journal} {\bibinfo  {journal} {JHEP}\ }\textbf {\bibinfo {volume} {03}},\
  \bibinfo {pages} {058}},\ \Eprint {https://arxiv.org/abs/1109.6305}
  {arXiv:1109.6305 [hep-th]} \BibitemShut {NoStop}%
\bibitem [{\citenamefont {Cavagli{\`a}}\ \emph {et~al.}(2014)\citenamefont
  {Cavagli{\`a}}, \citenamefont {Fioravanti}, \citenamefont {Gromov},\ and\
  \citenamefont {Tateo}}]{Cavaglia:2014exa}%
  \BibitemOpen
  \bibfield  {author} {\bibinfo {author} {\bibfnamefont {A.}~\bibnamefont
  {Cavagli{\`a}}}, \bibinfo {author} {\bibfnamefont {D.}~\bibnamefont
  {Fioravanti}}, \bibinfo {author} {\bibfnamefont {N.}~\bibnamefont {Gromov}},\
  and\ \bibinfo {author} {\bibfnamefont {R.}~\bibnamefont {Tateo}},\ }\href
  {https://doi.org/10.1103/PhysRevLett.113.021601} {\bibfield  {journal}
  {\bibinfo  {journal} {Phys. Rev. Lett.}\ }\textbf {\bibinfo {volume} {113}},\
  \bibinfo {pages} {021601} (\bibinfo {year} {2014})},\ \Eprint
  {https://arxiv.org/abs/1403.1859} {arXiv:1403.1859 [hep-th]} \BibitemShut
  {NoStop}%
\bibitem [{\citenamefont {Bombardelli}\ \emph {et~al.}(2017)\citenamefont
  {Bombardelli}, \citenamefont {Cavagli{\`a}}, \citenamefont {Fioravanti},
  \citenamefont {Gromov},\ and\ \citenamefont {Tateo}}]{Bombardelli:2017vhk}%
  \BibitemOpen
  \bibfield  {author} {\bibinfo {author} {\bibfnamefont {D.}~\bibnamefont
  {Bombardelli}}, \bibinfo {author} {\bibfnamefont {A.}~\bibnamefont
  {Cavagli{\`a}}}, \bibinfo {author} {\bibfnamefont {D.}~\bibnamefont
  {Fioravanti}}, \bibinfo {author} {\bibfnamefont {N.}~\bibnamefont {Gromov}},\
  and\ \bibinfo {author} {\bibfnamefont {R.}~\bibnamefont {Tateo}},\ }\href
  {https://doi.org/10.1007/JHEP09(2017)140} {\bibfield  {journal} {\bibinfo
  {journal} {JHEP}\ }\textbf {\bibinfo {volume} {09}},\ \bibinfo {pages}
  {140}},\ \Eprint {https://arxiv.org/abs/1701.00473} {arXiv:1701.00473
  [hep-th]} \BibitemShut {NoStop}%
\bibitem [{\citenamefont {Cavagli{\`a}}\ \emph
  {et~al.}(2021{\natexlab{b}})\citenamefont {Cavagli{\`a}}, \citenamefont
  {Gromov}, \citenamefont {Stefa{\'n}ski}, \citenamefont {Jr.},\ and\
  \citenamefont {Torrielli}}]{Cavaglia:2021eqr}%
  \BibitemOpen
  \bibfield  {author} {\bibinfo {author} {\bibfnamefont {A.}~\bibnamefont
  {Cavagli{\`a}}}, \bibinfo {author} {\bibfnamefont {N.}~\bibnamefont
  {Gromov}}, \bibinfo {author} {\bibfnamefont {B.}~\bibnamefont
  {Stefa{\'n}ski}, \bibfnamefont {Jr.}}, \bibinfo {author} {\bibnamefont
  {Jr.}},\ and\ \bibinfo {author} {\bibfnamefont {A.}~\bibnamefont
  {Torrielli}},\ }\href {https://doi.org/10.1007/JHEP12(2021)048} {\bibfield
  {journal} {\bibinfo  {journal} {JHEP}\ }\textbf {\bibinfo {volume} {12}},\
  \bibinfo {pages} {048}},\ \Eprint {https://arxiv.org/abs/2109.05500}
  {arXiv:2109.05500 [hep-th]} \BibitemShut {NoStop}%
\bibitem [{\citenamefont {Ekhammar}\ and\ \citenamefont
  {Volin}(2022)}]{Ekhammar:2021pys}%
  \BibitemOpen
  \bibfield  {author} {\bibinfo {author} {\bibfnamefont {S.}~\bibnamefont
  {Ekhammar}}\ and\ \bibinfo {author} {\bibfnamefont {D.}~\bibnamefont
  {Volin}},\ }\href {https://doi.org/10.1007/JHEP03(2022)192} {\bibfield
  {journal} {\bibinfo  {journal} {JHEP}\ }\textbf {\bibinfo {volume} {03}},\
  \bibinfo {pages} {192}},\ \Eprint {https://arxiv.org/abs/2109.06164}
  {arXiv:2109.06164 [math-ph]} \BibitemShut {NoStop}%
\bibitem [{\citenamefont {Gromov}(2025)}]{Wolfbook}%
  \BibitemOpen
  \bibfield  {author} {\bibinfo {author} {\bibfnamefont {N.}~\bibnamefont
  {Gromov}},\ }\href@noop {} {\bibinfo {title} {{Wolfbook: a VS Code extension
  for Wolfram Language notebooks}}},\ \bibinfo {howpublished}
  {\url{https://github.com/vanbaalon/wolfbook}} (\bibinfo {year} {2025}),\
  \bibinfo {note} {available on the VS Code Marketplace and Open
  VSX}\BibitemShut {NoStop}%
\end{thebibliography}%

\clearpage
\onecolumngrid
\appendix
\begin{center}
    \textbf{\large Supplemental Material}
\end{center}
\setcounter{equation}{0}
\setcounter{figure}{0}
\setcounter{table}{0}
\renewcommand{\theequation}{S\arabic{equation}}
\renewcommand{\thefigure}{S\arabic{figure}}
\renewcommand{\thetable}{S\arabic{table}}
\setcounter{secnumdepth}{2}  % numbered (lettered) sections in the SM so that \ref works
\counterwithout{equation}{section}  % one running S-numbering across the SM
\renewcommand{\theequation}{S\arabic{equation}}  % \counterwithout resets the format, so re-apply the prefix

\section{Numerical data}
\label{app:numdata}
Table~\ref{tab:S-finite} lists the spin-two dimension $\Delta$, the product $\cN\cA$ and the pole-free bottom quantities $\widehat C_{\cB}^{(p)}$ for $p=2,3,4$ at selected couplings between $g=0.1$ and $g=5$. Undoing the bottom normalisation defined in the main text gives $\cB_p$ and hence $C_p=\cN\cA\,\cB_p$.
\begin{table}[!ht]
  \caption{Finite-coupling data for the Konishi state: the spin-two dimension $\Delta$, the product $\cN\cA$ of normalisation and adjacent contributions, and the pole-free bottom quantities $\widehat C_{\cB}^{(p)}$ defined in the main text, for BPS dimensions $p=2,3,4$. These bottom quantities tend to one at strong coupling. Parentheses in the $\cN\cA$ column give the uncertainty in the last two digits.}
  \label{tab:S-finite}
  \begin{ruledtabular}
  \begin{tabular}{llllll}
    $g$ & $\Delta$ & $\cN\cA$ & $\widehat C_{\cB}^{(2)}$ & $\widehat C_{\cB}^{(3)}$ & $\widehat C_{\cB}^{(4)}$ \\
    \hline
    $0.1$ & $4.11550637795$ & $0.386157034641112473(31)$ & $0.390709291409$ & $0.244980653184$ & $0.166556427621$ \\
    $0.2$ & $4.41885988080$ & $0.339419698225031946(27)$ & $0.687874202394$ & $0.475804233675$ & $0.340970709709$ \\
    $0.3$ & $4.82694866228$ & $0.293256216848614275(45)$ & $0.861457312769$ & $0.659460405192$ & $0.505357961071$ \\
    $0.4$ & $5.27156518260$ & $0.254648071948140090(37)$ & $0.941920770603$ & $0.781258433754$ & $0.637609253892$ \\
    $0.5$ & $5.71272342479$ & $0.221996178403219909(18)$ & $0.973849931347$ & $0.853906923186$ & $0.732723428550$ \\
    $0.6$ & $6.13386281449$ & $0.1934642102289(29)$ & $0.986079049839$ & $0.896717783490$ & $0.798235692574$ \\
    $0.7$ & $6.53160607785$ & $0.168360446050(19)$ & $0.991177922975$ & $0.923124940259$ & $0.843524701786$ \\
    $0.8$ & $6.90750420602$ & $0.146414447489(17)$ & $0.993634197021$ & $0.940409393815$ & $0.875559138933$ \\
    $0.9$ & $7.26416958744$ & $0.1273639553026(48)$ & $0.994991354618$ & $0.952329638037$ & $0.898840339320$ \\
    $1.0$ & $7.60407071705$ & $0.1109000219616023647(89)$ & $0.995824839906$ & $0.960903313391$ & $0.916204272785$ \\
    $1.2$ & $8.24156344115$ & $0.084459964542(13)$ & $0.996770399510$ & $0.972159407155$ & $0.939792856738$ \\
    $1.4$ & $8.83269993932$ & $0.06478420929(87)$ & $0.997286249876$ & $0.979022356238$ & $0.954605685179$ \\
    $1.5$ & $9.11375404892$ & $0.05689838234653382(13)$ & $0.997466394305$ & $0.981493612970$ & $0.960014105869$ \\
    $1.6$ & $9.38631465637$ & $0.05006582425215029(71)$ & $0.997614763425$ & $0.983527909768$ & $0.964493084388$ \\
    $1.8$ & $9.90877170856$ & $0.038975880004321(21)$ & $0.997847517325$ & $0.986653545873$ & $0.971416103203$ \\
    $2.0$ & $10.4048217434$ & $0.030554502501(40)$ & $0.998025051267$ & $0.988916521696$ & $0.976452243130$ \\
    $2.2$ & $10.8781187975$ & $0.02410983993(19)$ & $0.998167580503$ & $0.990611676080$ & $0.980231487572$ \\
    $2.4$ & $11.3315440005$ & $0.01914108955(66)$ & $0.998286196881$ & $0.991917242518$ & $0.983141512756$ \\
    $2.5$ & $11.5515471110$ & $0.0170919633(11)$ & $0.998338705911$ & $0.992460960614$ & $0.984352068243$ \\
    $2.6$ & $11.7674165061$ & $0.0152832287(19)$ & $0.998387488656$ & $0.992946227754$ & $0.985431280653$ \\
    $2.8$ & $12.1876359167$ & $0.0122680111(44)$ & $0.998475638704$ & $0.993773168159$ & $0.987266530108$ \\
    $3.0$ & $12.5937814718$ & $0.009896548(51)$ & $0.998553455939$ & $0.994448887873$ & $0.988761038100$ \\
    $3.2$ & $12.9871829974$ & $0.008020792(32)$ & $0.998622917475$ & $0.995009039991$ & $0.989995007766$ \\
    $3.4$ & $13.3689728492$ & $0.006528970(26)$ & $0.998685470680$ & $0.995479255812$ & $0.991026295293$ \\
    $3.5$ & $13.5558230163$ & $0.005899736(24)$ & $0.998714509250$ & $0.995686724099$ & $0.991479701214$ \\
    $3.6$ & $13.7401247202$ & $0.005336445604(58)$ & $0.998742210287$ & $0.995878360256$ & $0.991897490978$ \\
    $3.8$ & $14.1014831562$ & $0.004378604635(84)$ & $0.998793987064$ & $0.996220442051$ & $0.992640514762$ \\
    $4.0$ & $14.4537863630$ & $0.0036057927(34)$ & $0.998841477255$ & $0.996516222874$ & $0.993279675005$ \\
    $4.2$ & $14.7976840778$ & $0.0029796267(39)$ & $0.998885228593$ & $0.996773984622$ & $0.993833750748$ \\
    $4.4$ & $15.1337517538$ & $0.0024702411(77)$ & $0.998925691777$ & $0.997000210890$ & $0.994317438099$ \\
    $4.5$ & $15.2990116925$ & $0.002251835(11)$ & $0.998944810215$ & $0.997103146742$ & $0.994536620354$ \\
    $4.6$ & $15.4625019450$ & $0.002054287(15)$ & $0.998963242648$ & $0.997200039251$ & $0.994742373375$ \\
    $4.8$ & $15.7843935400$ & $0.001713408(24)$ & $0.998998198211$ & $0.997377585665$ & $0.995117868862$ \\
    $5.0$ & $16.0998393215$ & $0.001433100(32)$ & $0.999030828478$ & $0.997536180831$ & $0.995451449911$ \\
  \end{tabular}
  \end{ruledtabular}
\end{table}

\end{document}